\documentclass[aip,jap,reprint,a4paper,floatfix,groupedaddress,citeautoscript]{revtex4-1}
\usepackage{mathptmx}
\usepackage{helvet}
\usepackage[utf8]{inputenc} 
\usepackage{graphicx}
\usepackage{epstopdf}
\usepackage{amsmath}
\usepackage{xspace}
\usepackage{sidecap}
\usepackage[
,textwidth=17.5cm
,textheight=23.5cm
,verbose
,dvips
]{geometry}

\begin{document}
\title{A physical model for the reverse leakage current in (In,Ga)N/GaN light-emitting diodes based on nanowires}

\author{M. Musolino}
\thanks{M. Musolino and D. van Treeck contributed equally to this work.}
\affiliation{Paul-Drude-Institut f\"{u}r Festk\"{o}rperelektronik,
Hausvogteiplatz 5--7, D-10117 Berlin, Germany}
\author{D. van Treeck}
\email[Author to whom correspondence should be addressed. Electronic mail: ]{treeck@pdi-berlin.de}
\affiliation{Paul-Drude-Institut f\"{u}r Festk\"{o}rperelektronik,
Hausvogteiplatz 5--7, D-10117 Berlin, Germany}
\author{A. Tahraoui}
\affiliation{Paul-Drude-Institut f\"{u}r Festk\"{o}rperelektronik,
Hausvogteiplatz 5--7, D-10117 Berlin, Germany}
\author{L. Scarparo}
\affiliation{Department of Information Engineering, University of Padova, Via Gradenigo 6/B, 35131 Padova, Italy}
\author{C. De Santi}
\affiliation{Department of Information Engineering, University of Padova, Via Gradenigo 6/B, 35131 Padova, Italy}
\author{M. Meneghini}
\affiliation{Department of Information Engineering, University of Padova, Via Gradenigo 6/B, 35131 Padova, Italy}
\author{E. Zanoni}
\affiliation{Department of Information Engineering, University of Padova, Via Gradenigo 6/B, 35131 Padova, Italy}
\author{L. Geelhaar}
\affiliation{Paul-Drude-Institut f\"{u}r Festk\"{o}rperelektronik,
Hausvogteiplatz 5--7, D-10117 Berlin, Germany}
\author{H. Riechert}
\affiliation{Paul-Drude-Institut f\"{u}r Festk\"{o}rperelektronik,
Hausvogteiplatz 5--7, D-10117 Berlin, Germany}


\begin{abstract} 

We investigated the origin of the high reverse leakage current in light emitting diodes (LEDs) based on (In,Ga)N/GaN nanowire (NW) ensembles grown by molecular beam epitaxy on Si substrates. To this end, capacitance deep level transient spectroscopy (DLTS) and temperature-dependent current-voltage (I-V) measurements were performed on a fully processed NW-LED. The DLTS measurements reveal the presence of two distinct electron traps with high concentrations in the depletion region of the p-i-n junction. These band gap states are located at energies of $570\pm20$ and $840\pm30$\,meV below the conduction band minimum. The physical origin of these deep level states is discussed. The temperature-dependent I-V characteristics, acquired between 83 and 403\,K, show that different conduction mechanisms cause the observed leakage current. On the basis of all these results, we developed a quantitative physical model for charge transport in the reverse bias regime. By taking into account the mutual interaction of variable range hopping and electron emission from Coulombic trap states, with the latter being described by phonon-assisted tunnelling and the Poole-Frenkel effect, we can model the experimental I-V curves in the entire range of temperatures with a consistent set of parameters. Our model should be applicable to planar GaN-based LEDs as well. Furthermore, possible approaches to decrease the leakage current in NW-LEDs are proposed.   

\end{abstract}

\maketitle

\section{Introduction}

III-N nanowires (NWs) are an attractive alternative to conventional planar layers as the basis for light-emitting diodes (LEDs) because they might pave the way to cost-effective phosphorless white LEDs~\cite{Li2012,Kolper2012}.
Despite the great effort invested by the scientific community into the fabrication of NW-based LEDs~\cite{Kikuchi2004,Sekiguchi2008,Lin2010,Guo2010,Bavencove2010,Armitage2010,Nguyen2011,Ra2013,Musolino2014}, so far, very little has been reported concerning the analysis of shallow traps and deep levels as well as their effect on charge carrier transport in these devices~\cite{Zhang2014e,Jahangir2014,Musolino2015b}. Band gap states significantly affect both the emission and the conduction properties in GaN based planar LEDs~\cite{Casey1996,Yan2010,Shan2011,Meneghini2014,AufderMaur2014}. In particular, the anomalous leakage current of LEDs, including subthreshold forward leakage current and reverse leakage current, is mainly due to the presence of deep states inside the band gap~\cite{Yan2010,Shan2011,AufderMaur2014}. Sizeable leakage current has been also observed in NW-based LEDs~\cite{Kikuchi2004,Lee2011a,Kishino2012,Tchernycheva2014,Musolino2015a}, but this issue has been addressed very rarely for this type of devices~\cite{Lee2011a}, and its physical origin is still unclear. In general, a reduction of the leakage current is a crucial aspect for improving the overall performance of LEDs, because leakage currents cause idle power consumption and affect the device reliability, luminescence efficiency as well as electrostatic discharge resilience~\cite{Cao2003c}.

In this work the origin of the reverse leakage current in a III-N NW-LED is carefully investigated. To this end, a fully processed device was characterized by capacitance-voltage measurements (C-V), capacitance deep level transient spectroscopy (DLTS) and temperature-dependent current-voltage (I-V) measurements. On the basis of these data, we have developed a quantitative physical model able to describe the experimental I-V curves of NW-LEDs in the reverse bias regime for a wide range of temperatures. The assumptions made in this study should remain valid also for planar devices based on III-N heterostructures, thus making our model applicable also to conventional planar LEDs.

\begin{figure}[b]
\includegraphics*[width=6.5cm]{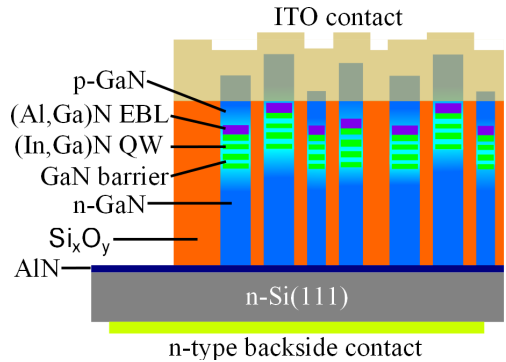} 
\caption{(Color online) Schematic of the employed NW-LED structure. Note that the various dimensions are not to scale.}
\label{LED}
\end{figure}

The NW-LED structure employed in this work was grown by molecular beam epitaxy (MBE) on an AlN-buffered n-doped Si(111) substrate with the help of self-assembly processes. The active region of the NW-LED consists of four axial (In,Ga)N quantum wells (QWs) with an average In content of 	approximately $25\%$, separated by three GaN barriers. The last QW is immediately followed by a Mg-doped (Al,Ga)N electron blocking layer (EBL). The active region is embedded between two doped GaN segments designed such that an n-i-p diode doping profile is created. A schematic of the structure is presented in figure~\ref{LED}. To fabricate the final devices, the NW ensemble was planarized by spin-on-glass, which fills the space between the NWs with amorphous Si$_{x}$O$_{y}$, and then a transparent indium-tin-oxide (ITO) p-type contact was sputtered on top of the sample. Finally, the n-type contact was created on the back side of the Si substrate. More details about the employed LED structure, the growth parameters as well as the fabrication process can be found in our previous publications~\cite{Musolino2014,Musolino2015a}.
 
\section{DLTS measurements}

An effective technique capable of probing the properties of electrically active states in the band gap of many semiconducting materials is capacitance DLTS. This technique measures the capacitance transients occurring in the space charge (depletion) region of a p-n junction when the charge carriers trapped by deep levels are released by means of thermally stimulated emission. A series of DLTS measurements was performed varying the emission rate window ($e_{n}$) from 2.4 to 2381\,$s^{-1}$, while the sample temperature was swept from 83 up to 360\,K. A quiescent reverse bias of -2\,V was employed; to ensure an adequate filling of the traps, filling pulses with amplitude of +1\,V and duration of 100\,ms were chosen. In this voltage range the boundary of the space charge region moves within the portion of the NW axis where the intrinsic segment of the p-i-n junction is located. Therefore, the DLTS technique can only probe the electrically active traps present inside or close to the active region of the NW-LED, which is located in the intrinsic segment of the junction. 

\begin{figure}[t]
\includegraphics*[width=8.5cm]{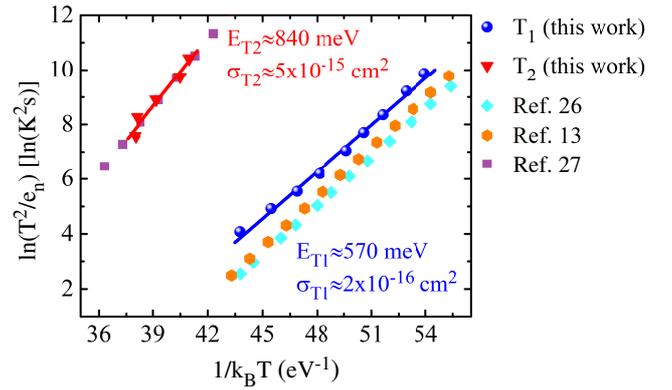} 
\caption{(Color online) Arrhenius plot for the deep levels T$_{1}$ (blue spheres) and T$_{2}$ (red triangles). In the graph, $T$ represents the temperature at which the DLTS peaks occur, while $e_n$ is the corresponding emission rate window. The slope and the intercept of the linear interpolations of the data (solid lines) yield the signature of the traps. For comparison, we plot also the data related to deep levels reported by Hierro \textit{et al.}~\cite{Hierro2001}, Jahangir \textit{et al.}~\cite{Jahangir2014}, and Osaka \textit{et al.}~\cite{Osaka2005} as cyan, purple, and orange symbols, respectively.}
\label{DLTS}
\end{figure}

A detailed analysis of the DLTS data is beyond the scope of this work and can be found elsewhere~\cite{Musolino2015b,Correction}; here, we summarize the outcome of this investigation. The DLTS measurements reveal the presence of two main electron traps, hereafter named T$_1$ and T$_2$. The signature of the deep levels was extracted from the Arrhenius plot shown in figure~\ref{DLTS}. From the slope of the linear interpolation of the data we obtained the energy of the traps measured from the conduction band edge ($E_{t}$), while from the intercept the apparent cross section ($\sigma$) was estimated. For trap $T_{1}$ we obtained $E_{T1}=(570\pm20)$\,meV and $\sigma_{T1}=(2\pm1)\times10^{-16}$\,cm$^{2}$; whereas for trap $T_{2}$ the extracted energy and cross section are $E_{T2}=(840\pm30)$\,meV and $\sigma_{T2}=(5\pm3)\times10^{-15}$\,cm$^{2}$. It is worth mentioning that DLTS measurements based on thermally stimulated emission from traps can only probe band gap states located within roughly 1\,eV from the conduction band minimum (or valence band maximum). In fact, deeper trap levels would require unrealistically high temperatures to be activated. This means that other deep levels, beyond those identified, might be present in the sample. 

To achieve a better understanding of the physical origin of the traps T$_{1}$ and T$_{2}$, we compare the obtained DLTS signatures with the ones reported in literature for deep levels in GaN based devices. The DLTS signature of the trap T$_{1}$ is in good agreement with the ones often found in n-type GaN layers grown by different techniques. For instance, the cyan symbols shown in figure~\ref{DLTS} refer to the trap level found by Hierro \textit{et. al}~\cite{Hierro2001} ($E_T\approx590$\,meV and $\sigma_T\approx2\times10^{-15}$\,cm$^{2}$). Although several energy values are reported in literature for this deep level (ranging from 520 to 600\,meV)~\cite{Hacke1994,Haase1996,Wang1998,Hierro2001,Umana-Membreno2007,Kindl2009,Peta2013}, its origin has often been attributed to nitrogen antisite defects ($N_{Ga}$).
Trap T$_{1}$ exhibits also a good agreement with the signature of the deep level recently reported for GaN NW based n$^+$-p junctions by Jahangir \textit{et al.}~\cite{Jahangir2014} and depicted by the orange hexagons in figure~\ref{DLTS} ($E_T\approx607$\,meV and $\sigma_T\approx3.3\times10^{-15}$\,cm$^{2}$). This trap state was attributed to the formation of chains of boundary dislocations (BDs) at the sidewalls of coalescing NWs. Both the deep levels reported by Hierro \textit{et al.} and Jahangir \textit{et al.} may likely be present in the investigated NW-LED. In fact, the formation of $N_{Ga}$ could be promoted by the highly N-rich ambient employed for the growth of NWs, while dislocations at the boundaries of coalesced NWs are likely also present in the studied sample~\cite{Consonni2009,Grossklaus2013a}. Hence, it seems to be rather plausible that the trap level T$_{1}$ originates from the coexistence of these two types of defects in the NW-LED, namely $N_{Ga}$ and chains of BDs. The former are likely located in the core of the NWs, whereas the latter form on the sidewalls. 

The signature of trap T$_{2}$ is similar to the ones already found in GaN planar layers grown by MBE ($E_{T}\approx890$\,meV~\cite{Peta2013} and $E_{T}\approx910$\,meV~\cite{Hierro2001}), by metalorganic vapor phase epitaxy ($E_{T}\approx800$\,meV~\cite{Kindl2009}), and also by hydride vapor phase epitaxy ($E_{T}\approx808$\,meV~\cite{Osaka2005}). As an example, in figure~\ref{DLTS} we plot the signature of the trap found by Osaka \textit{et al.}~\cite{Osaka2005} (purple squares); the overlap with the data obtained for trap T$_{2}$ is very good. This trap level has been attributed to interacting point defects arranged along threading dislocations (TDs)~\cite{Hierro2001,Kindl2009}. The NW-LEDs are free of extended defects such as TDs; nevertheless, dangling bonds similar to those present in threading dislocations might exist on the surface of the NWs. Therefore, we tentatively assign the trap level T$_{2}$ to defect complexes (likely dangling bonds) located on the sidewalls of the NWs. 

Assuming that the defects are uniformly distributed inside the volume of the depletion region, the concentration of traps $N_T$ can be extrapolated by means of the formula~\cite{Lang1974}
\begin{align}
\label{eq:DLTS_dens}
N_{T}=2\frac{\Delta C}{C}N_{CV}  
\end{align}
where $N_{CV}$ is the net charge density in the active region (extrapolated by means of C-V measurements); whereas the ratio $\Delta C/C$ represents the normalized DLTS signal. The obtained trap concentrations are $N_{T1}\approx5\times10^{16}$\,cm$^{-3}$ and $N_{T2}\approx1\times10^{16}$\,cm$^{-3}$ for trap T$_{1}$ and T$_{2}$, respectively. These values are between one and two orders of magnitude higher than the ones reported for similar defects observed in planar (In,Ga)N/GaN blue LEDs~\cite{Meneghini2014} (\textit{i.\,e.}, $N_{T}\approx 3\--27\times10^{14}$\,cm$^{-3}$) and NW based n$^{+}$-p junctions~\cite{Jahangir2014} (\textit{i.\,e.}, $N_{T}\approx 2\--48\times10^{14}$\,cm$^{-3}$) grown by MOVPE (metalorganic vapor phase epitaxy) and MBE, respectively. This comparison would indicate that the high values of the defect density found in our green NW-LEDs might be caused by two main factors: the presence of an (In,Ga)N alloy with relatively high In content and the use of growth temperatures that are significantly lower than the ones commonly employed in MOVPE. 


\section{Charge transport in reverse bias regime}

To further elucidate the nature of the involved current conduction processes, temperature-dependent I-V measurements were carried out. Such a study can help to reveal the nature of the involved conduction mechanisms, which are often characterized by different temperature dependences. Before we present the experimental data in the next section, we discuss the transport mechanisms that can be expected to be relevant. As the basis for the following analysis, we have to make some assumptions. First, the contribution of the drift-diffusion and Sah-Noyce-Shockley generation-recombination process to the reverse current is considered to be negligible. This assumption is plausible for wide band gap semiconductors such as GaN~\cite{SzeBook}. Second, the insulating Si$_{x}$O$_{y}$ matrix used to planarize the NW ensemble does not provide further leakage paths. This assumption is supported by conductivity measurements performed on a separate sample with a roughly 500-nm-thick Si$_{x}$O$_{y}$ layer deposited on Si without any NWs; the current observed across the insulating layer was negligible, namely few pico-amps. Under these assumptions, it seems reasonable to say that the leakage paths are located within the p-i-n junction of the NW-LEDs or along the side-walls of the NWs. 

\begin{figure}[t]
\includegraphics*[width=8cm]{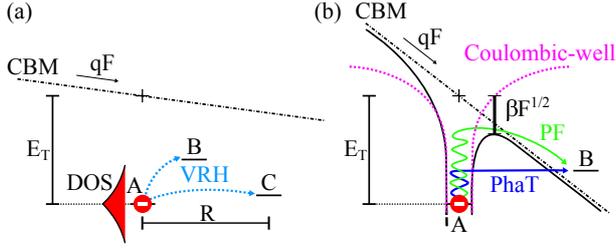} 
\caption{(Color online) Energy diagram of localized trap states in the band gap with a schematic description of the conduction mechanisms. The trap A located at the energy $E_{T}$ below the conduction band minimum (CBM) is the supply centre and is negatively charged. Its DOS exhibits an exponential distribution. (a) VRH process: the electrons hop from a trap A to an empty state located at a distance R (\textit{i.\,e.}, B or C). (b) PF effect (green) and PhaT (blue); the trap state is described as a Coulombic-well, and the height of the barrier is decreased by a factor $\beta F^{1/2}$.}
\label{Theory}
\end{figure}

The space charge region along the axis of the NW-LEDs can be imagined in reverse bias as a semi-insulating segment under a strong electric field characterized by the presence of a high density of electron traps, as revealed by the DLTS measurements. Our analysis of current conduction has been inspired by similar studies on GaN layers~\cite{Look1996} and planar (In,Ga)N/GaN LEDs~\cite{Shan2011}. These reports build on a current conduction model for disordered semi-insulating material proposed by N. F. Mott~\cite{Mott1969} and R. M. Hill~\cite{Hill1971}. According to this theory, the current conduction process is due to the hopping of electrons from one trap state to another one. The model assumes that many trap states are present at different energies ($E_{T}$) inside the band gap of the semiconductor, and that these traps are characterized by a density of states (DOS) with an exponential energy distribution: DOS$=D_{T}\exp(-|E-E_{T}|/U)$. In this equation $D_{T}$ is the trap density of states at the energy of the supply trap in units of volume$^{-1}$ energy$^{-1}$, whereas $U$ represents the characteristic energy of the distribution; it may also be seen as the standard deviation of the distribution, so that about the 93\% of the available states are distributed inside an energy range equal to $2.62\times U$. A schematic description of this conduction process is depicted in figure~\ref{Theory}(a). Electrons can hop from a supply trap [labelled A in figure~\ref{Theory}(a)], to an empty one located at a distance R [for instance trap B or C in figure~\ref{Theory}(a)] and inside an energy range around $E_T$; hereinafter named variable range hopping (VRH). The hopping mechanism is enhanced in the presence of an electric field ($F$), which facilitates the hop from one trap state to the next one. 
Hill obtained the following expression for the current ($I_{VRH}$) due to the VRH conduction process~\cite{Hill1971}:
\begin{align}
\label{eq:VRH}
I_{VRH}=I_{0}^{VRH}\exp \left[-1.76\left(\frac{T_{0}}{T}\right)^{1/4}+C_{VRH}\left(\frac{T_{0}}{T}\right)^{3/4}F^{2}\right]  
\end{align}
In this equation, the pre-exponential term $I_{0}^{VRH}$ converts the emission rate of the trap state into current,
$T_{0}$ is a characteristic temperature parameter defined as $T_{0}= 18/(k_{B}D_{T}a^{3})$ where $a$ is the localization radius of the wave function describing the trapped electron, and $k_{B}$ represents the Boltzmann constant. The parameter $C_{VRH}$ groups together several quantities, and it is defined as $C_{VRH}=4.626\times 10^{-3}(qa/U)^{2}$ where $q$ represents the elementary charge. Note that $a$ and $U$ are in principle unknown constants, which describe physical properties related to the intimate nature of the trap states. Equation~(\ref{eq:VRH}) is valid for moderate electric fields, namely $qFa\lesssim 5k_{B}T$. For T=300\,K and assuming a localization radius of the wave function equal to the Bohr radius of an electron bound to an oxygen donor in GaN, namely $a=2.8$\,nm~\cite{Echeverria2008}, we obtain $F\lesssim 5\times10^{5}$\,V/cm. For electric fields higher than this value hopping should not be the dominant current conduction mechanism any longer. 

At high electric fields the potential surrounding the trap state, which is usually assumed to be Coulombic, is strongly deformed: the height of the Coulomb barrier is lowered and its width is thinned. The emission of charge carriers resulting from the lowering of the Coulomb potential barrier can be described by the Poole-Frenkel (PF) effect, while the tunnelling through the thinned barrier is usually modelled by means of phonon-assisted tunnelling (PhaT). Figure~\ref{Theory}(b) shows a graphical representation of the charge carrier emission from a Coulomb state in the high-field case. At zero temperature only direct tunnelling into the conduction band at the energy level of the Coulomb centre is possible. With increasing temperature, due to electron-phonon coupling, the electrons gain energy and can thus tunnel through the potential barrier also towards states at higher energy (\textit{i.\,e.}, state B). The electrons can either tunnel through the barrier (PhaT) or overpass it (PF effect). At high temperatures and strong electric field the PF effect should dominate over PhaT.
The contribution to the electron emission due to the PF effect can be described by the formula~\cite{Frenkel1938,Hill1971b} 
\begin{align}
\label{eq:PF}
e_{PF}\propto\exp\left(-\frac{E_{T}-\beta F^{1/2}}{k_{B}T}\right)
\end{align}
where $e_{PF}$ represents the electron emission rate from donor states located at the energy $E_{T}$ below the conduction band edge. The term $\beta F^{1/2}$ describes the lowering of the Coulomb barrier; $\beta=(Zq^{3}/\pi\epsilon_{r}\epsilon_{0})^{1/2}$ is called Poole-Frenkel coefficient. In the definition of $\beta$, $Z$ is the charge state of the Coulomb centre, whereas $\epsilon_{0}$ and $\epsilon_{r}$ are the vacuum and relative permittivity, respectively.

An analytical expression of the electron emission rate due to PhaT was obtained by Vincent \textit{et al.}~\cite{Vincent1979}: 
\begin{eqnarray}
\label{eq:PhaT}
e_{PhaT}\propto\exp\left(-\frac{E_{T}}{k_{B}T}\right)\times\int_{\beta F^{1/2}/k_{B}T}^{E_{T}/k_{B}T}\exp{\Bigg\lbrace} z-z^{3/2}\nonumber\\
\times\left(\frac{4}{3}\frac{(m_{e}^{*})^{1/2}(k_{B}T)^{3/2}}{q\hbar F}\right)\left[1-\left(\frac{\beta F^{1/2}}{zk_{B}T}\right)\right]{\Bigg\rbrace} dz
\end{eqnarray}
where the integral sums over the whole depth of the Coulombic well, $m_{e}^{*}$ is the effective mass of the electron and $\hbar$ the Planck constant divided by $2\pi$. The total current emitted by Coulombic trap states ($I_{CT}$) has to take into account both contributions, it is thus proportional to the sum of $e_{PF}$ and $e_{PhaT}$~\cite{Vincent1979}  
\begin{align}
\label{eq:CT}
I_{CT}=I_{0}^{CT}\left(e_{PF}+e_{PhaT}\right)
\end{align}
where the factor $I_{0}^{CT}$ converts the emission rate into current, it is proportional to the density of traps ($N_{T}$) which are actually contributing to the conduction process. 

In the next sections we try to fit the experimental I-V-T curves with equations~(\ref{eq:VRH}) and (\ref{eq:CT}) to verify whether the introduced conduction mechanisms can describe the transport properties of the NW-LEDs. 

\section{Analysis of the I-T characteristics}

\begin{figure}[t]
\includegraphics*[width=8cm]{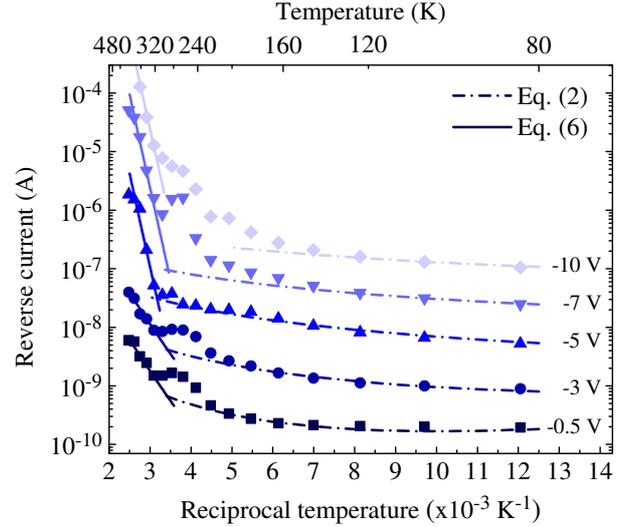} 
\caption{(Color online) I-T characteristics of the NW-LED measured at five different reverse biases for temperatures ranging from 83 to 403\,K (symbols). The data were fitted using equations~(\ref{eq:VRH}) and (\ref{eq:CT_T}), indicated by the dashed-dotted and solid lines, respectively.}
\label{I-T}
\end{figure}

The leakage current in the reverse bias regime was investigated by means of temperature-dependent I-V measurements for temperatures ranging from 83 to 403\,K. The data were acquired in a dark ambient by means of a cold finger cryostat filled with liquid nitrogen and a semiconductor parameter analyser, model HP 4155. Figure~\ref{I-T} shows the evolution of the reverse current with the reciprocal temperature, $1/T$, for five different applied biases. In the graph two main regions are clearly identifiable. For temperatures lower than approximately 240\,K (namely, $1/T>4.16\times10^{-3}$\,K$^{-1}$), the current varies slowly with temperature, while above 240\,K a rather fast change in the slopes of the curves is visible. This behaviour suggests the presence of at least two different conduction channels, one of which is thermally activated.

The current-temperature (I-T) curves were fitted by means of equation~(\ref{eq:VRH}), using as fitting parameters $I_{0}^{VRH}$ and $T_{0}$, and keeping $C_{VRH}$ constant and equal to $2\times 10^{-18}$\,C$^{2}$\,m$^{2}$\,eV$^{-2}$. A detailed discussion of the role and the meaning of the fitting parameters is presented section~\ref{sec:Par}. The VRH model, depicted by the dashed-dotted lines in figure~\ref{I-T}, can well describe the temperature dependence of the reverse current only for $T\lesssim240$\,K and biases ranging from -0.5 to -5\,V. For higher negative biases, corresponding to higher electric fields in the depletion region, the discrepancy between experimental points and VRH model becomes larger and larger. The value of $T_{0}$ obtained from the fit of the data related to a bias of -5\,V, namely $1\times10^{5}$\,K, is about one order of magnitude lower than the ones reported by Shan \textit{et al.}~\cite{Shan2011} and Jung \textit{et al.}~\cite{Jung2014} for planar (In,Ga)N/GaN LEDs. This would indicate the presence of a higher density of traps, $N_{T}$, in the NW-LEDs than in the planar counterparts (in fact, $N_{T}\propto D_{T}\propto 1/T_{0}$), thus supporting the outcome obtained by the DLTS measurements. 

Above 240\,K a rapid increase of the current with temperature is observed. This behaviour seems to suggest the appearance of a thermally activated emission process different from VRH. The temperature dependence of the current emitted by trap states with thermal activation energy $E_{a}$ can be written as~\cite{Mazzola1994}
\begin{align}
\label{eq:CT_T}
I_{T}=AT^{2}\exp\left(-\frac{E_{a}}{k_{B}T}\right)
\end{align}
where $A$ is a proportionality factor which produces an almost rigid vertical shift of the curve, whereas $E_{a}$ defines its slope. It is worth noting that the temperature dependence defined by equation~(\ref{eq:CT_T}) coincides with the main temperature dependence of the current emitted by Coulombic traps, see equations~(\ref{eq:PF})$\--$(\ref{eq:CT}). The fit of the data is represented by the solid lines in figure~\ref{I-T}; the evolution of the current is qualitatively described by equation~(\ref{eq:CT_T}) for temperatures higher than 300\,K. Two different slopes of the curves are obtained, corresponding to two different activation energies: $E_{a}\approx 150$\,meV for $|V|\lesssim 3$\,V and $E_{a}\approx 570$\,meV for $|V|\gtrsim 5$\,V. Interestingly, the latter activation energy is equal to the one obtained by means of DLTS for trap T$_1$; therefore, these defects are most likely responsible for the current conduction properties of the NW-LEDs at temperatures and absolute biases higher than 240\,K and -5\,V, respectively. The energy $E_{a}\approx 150$\,meV might be related to the activation of shallow traps close to the valence band; however, its actual origin remains to be understood. The low bias regime is further discussed in section~\ref{sec:Par}.  

The analysis of the I-T curves thus suggests that the current conduction is dominated by VRH at low temperatures and small reverse biases ($T\lesssim 240$\,K and $|V|\lesssim 5$\,V). At high temperatures and biases ($T\gtrsim 300$\,K and $|V|\gtrsim 5$\,V) the dominating transport mechanism is likely due to thermally activated emission from deep states located 570\,meV below the conduction band edge in the depletion region of the NW-LED.

\section{Analysis of the I-V characteristics}

\begin{figure}[b]
\includegraphics*[width=7cm]{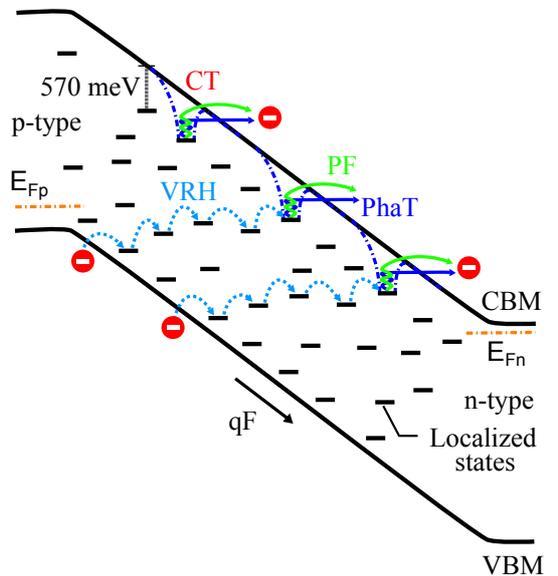} 
\caption{(Color online) The schematic depicts the depletion region of the p-i-n junction under reverse bias, for sake of simplicity the (In,Ga)N QWs are omitted. The electrons hop (VRH) from the VBM up to the CT states located 570\,meV below the CBM. Then, the electrons are emitted by means of the PF effect and PhaT.}
\label{Model}
\end{figure}

Both I-T and DLTS measurements support the idea that deep states located 570\,meV below the CBM significantly contribute to the current transport in reversely biased NW-LEDs. However, the mechanism by which the charge carriers overcome the energy gap present between the valence band maximum (VBM) and these trap states located in the vicinity of the CBM has still to be understood. A possible explanation could be that electrons from the VBM of p-type GaN tunnel into shallow states, and then move towards the other side of the band gap by multiple hops. A schematic description of this conduction model is depicted in figure~\ref{Model}. A similar idea has already been employed to describe the reverse leakage current in planar (In,Ga)N/GaN LEDs~\cite{Shan2011}. 

In order to elucidate the role of the different transport mechanisms in details, we develop in the following a physical model that allows the quantitative description of I-V curves over a large range of reverse biases and temperatures. In our model, we assume that the electron hopping from the VBM up to the trap states located 570\,meV below the CBM is well described by the VRH process. Already few tens of defects in the depletion region of the NW-LEDs (corresponding to densities in the range $10^{16}\--10^{17}$\,cm$^{-3}$) can enable the VRH conduction across the band gap. In fact, in the investigated sample the average hopping distance [defined as $\langle R\rangle=a(T_{0}/T)^{1/4}$]~\cite{Shklovskii-Book} is of the order of tens of nanometres, thus making possible electron flow across the entire depletion region with few hops. For T=200\,K and assuming $a=2.8$\,nm, and $T_{0}=5\times10^{5}$\,K, we obtain $\langle R\rangle\approx20$\,nm. Once the electrons reach the vicinity of the CBM, they are trapped by the Coulombic potential surrounding the deep states, and then emitted by means of a thermally activated process, namely PhaT and the PF effect. 

In general, each charged trap state could emit electrons by means of VRH, PhaT, and the PF effect. Nevertheless, the last two mechanisms are unlikely to occur in deep levels located far from the CBM (or from the VBM) because of the high energy barrier that the electron has to overcome; in these cases it is more likely for an electron to hop to the next localized state. In contrast, in trap states close to the CBM the current transport should be dominated by PhaT and the PF effect rather than VRH; in fact, in these cases the emission of electrons into the continuum states in the conduction band should be more favourable than hopping. In this picture of the system, the VRH mechanism acts as charge carrier supplier for the Coulomb trap (CT) states close to the CBM (see figure~\ref{Model}). Therefore, the VRH conduction channel can be considered in series to the CT system described by PhaT and the PF effect. The total electron leakage current across the band gap ($I_{tot}$) can be expressed by the analytical relation 
\begin{align}
\label{eq:Serie}
I_{tot}\approx\frac{I_{VRH}I_{CT}}{I_{VRH}+I_{CT}}
\end{align} 
where $I_{VRH}$ and $I_{CT}$ are defined by equations~(\ref{eq:VRH}) and (\ref{eq:CT}), respectively. This formula describes the competition between two mechanisms that limit each other, such that the slower process governs the faster one. Shockley and Read found that a similar relation describes the generation-recombination processes in trap states when the mechanism that fills the traps is in competition with the one that empties them~\cite{Shockley1952}. 

\begin{figure}[b]
\includegraphics*[width=8.5cm]{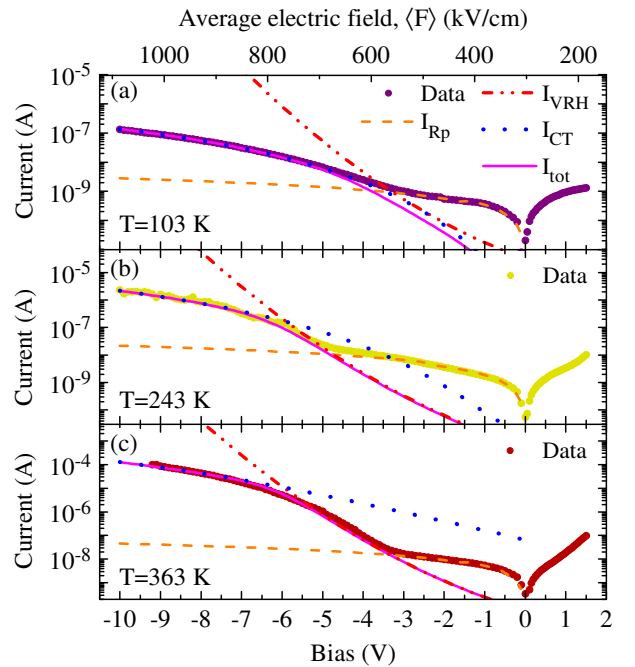} 
\caption{(Color online) Experimental I-V curves in reverse bias (data points) plotted in semi-logarithmic scale and contributions of each conduction channel to the total current plotted for three different temperatures: (a) 103\,K, (b) 243\,K, and (c) 363\,K.}
\label{I-V}
\end{figure}

In order to illustrate the success of our model and the key insights we have obtained on its basis, we present initially only a representative subset of the experimental I-V curves that exhibit the most characteristic features. Figures~\ref{I-V} (a)$\--$(c) show the evolution of the reverse current with the applied bias for three different temperatures: 103, 243, and 363\,K. The experimental curves exhibit a peculiar hump-like shape. For small reverse voltages, approximately between 0 and -3\,V, the leakage current has a nearly linear behaviour (it appears as a logarithmic curve in semi-logarithmic scale, see the orange dashed lines). In contrast, the current increases super-linearly as the reverse voltage is swept from -4 to -10\,V. As already discussed in the previous section, the reverse current exhibits also a clear dependence on the temperature, increasing much faster for $T\gtrsim 240$\,K than for low temperatures, compare figures~\ref{I-V} (a) and (b). This behaviour is due to the presence of a thermally activated transport mechanism, which dominates the current conduction for temperatures higher than about 240\,K. 

The current-voltage (I-V) curves were fitted by means of equation~(\ref{eq:Serie}). The average electric field $\left(\langle F\rangle\right)$ in the depletion region was obtained for each applied bias from C-V measurements (data not shown here), and used in place of $F$ in equations~(\ref{eq:VRH}) and (\ref{eq:CT}). The choice of the fitting parameters and their meaning are discussed in detail in the next section. The resulting fits to the data obtained by means of equation~(\ref{eq:Serie}) are shown by the magenta solid lines in figures~\ref{I-V} (a)$\--$(c). The different contributions, namely $I_{VRH}$ and $I_{CT}$, are depicted by the red dashed-dotted and blue dotted lines, respectively. At low temperature (T=103\,K), VRH would potentially supply much more current than the one that the CT states are able to emit (compare red and blue curves). The total current is thus limited by the low emission rate of the CT states. For negative biases higher than about -4\,V, the data are perfectly fitted by $I_{CT}$. At intermediate temperature (T=243\,K), the thermally activated emission rate of the CT states significantly increases, and the charge carriers supplied by VRH for biases between 0 and -6\,V are not able to follow the fast emission rate of the CT states any longer. In this bias range the low emission rate of the VRH process limits the current emitted by the CT states. In contrast, for reverse voltages higher than -6\,V the emission rate of the VRH process overtakes the one of the CT, and the total current is again limited by the emission from CT states. At high temperature (T=363\,K), the evolution of the I-V curves is similar to the one observed at T=243\,K. A quantitative comparison of the two contributions describing the electron emission from the CT state reveals that PhaT dominates over the PF effect for all the considered temperatures.

The mutual interaction of VRH and emission from CT states describes fairly well the I-V curves for $|V|\gtrsim 3$\,V; nevertheless, in the low bias (low field) regime neither equation~(\ref{eq:VRH}) nor equation~(\ref{eq:CT}) can describe the voltage dependence of the measured current. In order to fit the experimental data points in the low bias regime, another term with linear dependence on the voltage (called $I_{R_{P}}$) was added to equation~(\ref{eq:Serie}). Hence, the total reverse current becomes 
\begin{align}
\label{eq:Itot}
I_{tot}'=I_{tot}+I_{R_{P}}=\frac{I_{VRH}I_{CT}}{I_{VRH}+I_{CT}}+\frac{V}{R_{P}}
\end{align}
where $R_{P}$, named parallel resistance, is an empirical parameter that takes into account further parallel leakage mechanisms not included in our model. The physical origin of this Ohmic-like contribution is unclear. It might be related to additional leakage current paths present on the surface of the NWs or to other types of band-to-band tunnelling not considered in this study; for example, similar to the one responsible for the low forward bias current conduction in (In,Ga)N/GaN LEDs~\cite{AufderMaur2014}. Another possible explanation could be a weaker field dependence of the VRH at low electric fields, as predicted by Hill~\cite{Hill1971}. The contribution of the linear term defined by $I_{R_{P}}$ is depicted by the orange dashed lines in figure~\ref{I-V}; it describes very well the experimental data points for $|V|\lesssim 3$\,V.

\begin{figure}[t]
\includegraphics*[width=8cm]{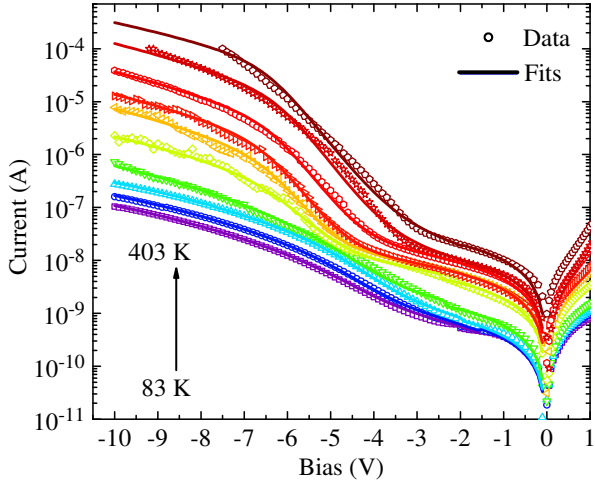} 
\caption{(Color online) Experimental I-V curves in reverse bias (data points) and fits (solid lines) with equation~(\ref{eq:Itot}) plotted in semi-logarithmic scale. The curves correspond to temperatures of 83, 123, 163, 203, 243, 303, 323, 343, 363, and 403\,K (from bottom to top).}
\label{I-V-T}
\end{figure}

The final fits of the I-V curves with equation~(\ref{eq:Itot}) are shown in figure~\ref{I-V-T} for different temperatures ranging from 83 to 403\,K. The agreement with the data points is excellent in the entire bias range and for all the temperatures up to 323\,K. For higher temperatures some discrepancy occurs, but the qualitative behaviour is still well described by the employed model. From the analysis of the data we can conclude that the peculiar hump-like shape observed in the I-V curves in reverse bias is likely due to the fast increase of the emission rate of the CT states with temperature.

\section{Discussion of the fitting parameters}\label{sec:Par}

Both I-T and I-V curves were fitted by means of equation~(\ref{eq:VRH}), which is characterized by five quantities: $I_{0}^{VRH}$, $T_{0}$, $C_{VRH}$, $T$, and $F$. The latter two ($T$ and $F$) are defined by the measurement conditions. The parameters $C_{VRH}$ and $T_0$ define the steepness of the curves; however, because of their different weight in equation~(\ref{eq:VRH}), they influence the curvature of the plots differently. By means of a process of iterative optimization, see Appendix, it was possible to minimize the discrepancy between the experimental I-V data and the fit. The operation was repeated for different values of $C_{VRH}$ and for various temperatures. The optimum curvature and, hence, the best fits to the data were obtained for $C_{VRH}\approx 2\times 10^{-18}$\,m$^{2}$eV$^{-2}$ in a wide range of temperatures. Thus, this value of $C_{VRH}$ was kept constant for all the fits. Assuming a localization radius ($a$) of the wave function describing the trapped charge carriers equal to the Bohr radius of an electron bound to an oxygen donor in GaN, namely $a\equiv a_{B}=2.8$\,nm~\cite{Echeverria2008}, we obtain $U\approx 135$\,meV, which is a reasonable value for the scale factor of the DOS distribution. The two remaining parameters, $I_{0}^{VRH}$ and $T_{0}$, were used to fit the experimental data; $T_{0}$ controls the slope of the curves, whereas a variation of $I_{0}^{VRH}$ produces an almost rigid vertical shift of the VRH current. 

The emission from the Coulomb trap states described by equation~(\ref{eq:CT}) is characterized by six quantities: $I_{0}^{CT}$, $E_{T}$, $\beta$, $m_{e}^{*}$, $T$, and $F$. Again, the latter two are defined by the measurement conditions. In our model we assumed that the CT states have the same nature as the trap $T_{1}$ observed by DLTS (\textit{i.\,e.}, $N_{Ga}$ defects), the parameter $E_{T}$ was thus chosen equal to $E_{T1}=570$\,meV. Assuming doubly ionized Coulombic centres (\textit{i.\,e.}, $Z=2$), as expected for $N_{Ga}$ defects in GaN~\cite{Mattila1996}, and using the static relative permittivity of GaN $\epsilon_{r}=9.7$, we obtained $\beta=3.45\times10^{-5}$\,eVV$^{-1/2}$m$^{1/2}$. The two remaining parameters, $I_{0}^{CT}$ and $m_{e}^{*}$, were used to fit the experimental data; $m_{e}^{*}$ changes the slope of the curves, whereas a variation of $I_{0}^{CT}$ produces an almost rigid vertical shift. Finally, the reverse current at low negative biases was described by means of the empirical parameter $R_{P}$. 

\begin{table}[t]
\caption[]{\label{tab:Par_rev_IV}List of the parameters obtained by fitting the I-V curves for temperatures ranging from 83 to 403\,K.}
\begin{ruledtabular}
\begin{tabular}{c|cc|cc|cc}
\multicolumn{5}{c}{\qquad\qquad\quad\quad VRH \qquad\qquad\quad  Emission from CT} \\ \hline
T (K) & $I_{0}^{VRH}$ (mA) & $T_{0}$ ($\times10^{6}$\,K) & $I_{0}^{CT}$ (A) & $m^{*}_{e}/m_{e}$ & $R_{p}$ (G$\Omega$)\\
83 & 0.06 & 1.0 & $2.4\times10^{-6}$ & 0.10 & 3.40 \\
103 & 0.07 & 1.3 & $3.0\times10^{-6}$ & 0.10 & 3.25 \\
123 & 0.08 & 1.5 & $3.5\times10^{-6}$ & 0.10 & 3.25 \\
143 & 0.10 & 1.8 & $4.4\times10^{-6}$ & 0.10 & 2.95 \\
163 & 0.13 & 2.1 & $5.8\times10^{-6}$ & 0.10 & 2.70 \\
183 & 0.15 & 2.5 & $9.1\times10^{-6}$ & 0.11 & 2.15 \\
203 & 0.14 & 2.8 & $2.1\times10^{-5}$ & 0.17 & 1.60 \\
223 & 0.15 & 3.0 & $3.3\times10^{-5}$ & 0.17 & 1.15 \\
243 & 0.05 & 2.6 & $6.7\times10^{-5}$ & 0.15 & 0.50 \\
263 & 0.03 & 3.4 & $4.3\times10^{-5}$ & 0.05 & 0.37 \\
283 & 0.03 & 3.9 & $1.3\times10^{-4}$ & 0.12 & 0.32 \\
303 & 0.03 & 3.8 & $3.5\times10^{-4}$ & 0.22 & 0.35 \\
323 & 0.06 & 4.0 & $6.0\times10^{-4}$ & 0.22 & 0.36 \\
343 & 0.20 & 4.3 & $1.8\times10^{-3}$ & 0.25 & 0.23 \\
363 & 1.00 & 4.6 & $5.9\times10^{-3}$ & 0.26 & 0.21 \\
383 & 2.80 & 4.4 & $7.5\times10^{-3}$ & 0.20 & 0.13\\
403 & 3.00 & 4.7 & $8.5\times10^{-3}$ & 0.20 & 0.12 \\
\end{tabular}
\end{ruledtabular}
\end{table}

All in all, we were able to model both the voltage and the temperature dependence of the reverse leakage current observed in NW-LEDs by means of only five parameters. Since these parameters are independent from each other, they can be unequivocally determined within a certain inaccuracy, which we estimate to be smaller than a few percent (see Appendix). The values employed to fit the I-V curves are listed in table~\ref{tab:Par_rev_IV} and are discussed below. 
 
\begin{figure}[t]
\includegraphics*[width=8cm]{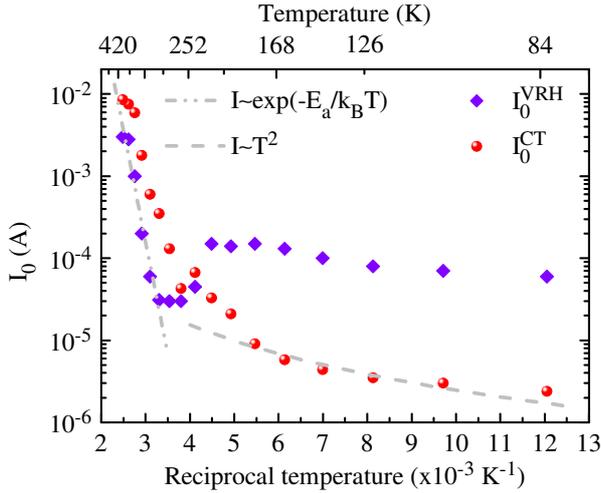} 
\caption{(Color online) Arrhenius plot of the pre-exponential terms; the violet diamond and the red circle points represent the parameters $I_{0}^{VRH}$ and $I_{0}^{CT}$, respectively. The grey dotted-dashed line is a linear interpolation of the data at high temperatures, whereas the grey dashed curve depicts a hypothetical quadratic temperature dependence of the current ($I\sim T^{2}$).}
\label{Io}
\end{figure}

Figure~\ref{Io} shows the variation of $I_{0}^{VRH}$ and $I_{0}^{CT}$ with reciprocal temperature. The parameter $I_{0}^{VRH}$ exhibits small fluctuations for temperatures ranging from 83 up to 323\,K, but not a well defined behaviour. Indeed, no particular temperature dependence of the pre-exponential factor is expected from the VRH model~\cite{Hill1971}. For temperatures higher than 323\,K an exponential increase of $I_{0}^{VRH}$ is observed. The slope of the Arrhenius plot (see grey dotted-dashed line) indicates a thermal activation energy ($E_{a}$) of this process roughly equal to 540\,meV. Interestingly, this value is very close to the energy of the Coulomb traps obtained from the analysis of both DLTS and I-T measurements. 
Such a strong temperature dependence of the pre-exponential factors would suggest that for temperatures higher than about 300\,K the charge carriers transport between deep level states is no longer described by VRH [\textit{i.\,e.}, equation~(\ref{eq:VRH})]. Indeed, this conduction mechanism is supposed to dominate only at low temperatures, when the hop into remote trap states with energy close to $E_{T}$ is more favourable than the hop into neighbouring states with energies very different than that of the supply trap. In contrast, at high temperatures the hopping into close states with energies higher than $E_{T}$ is favoured; this transport mechanism is usually referred to as nearest-neighbour hopping, and is characterized by a temperature dependence stronger than that of VRH~\cite{Shklovskii-Book}. At temperatures lower than about 200\,K, the pre-exponential factor $I_{0}^{CT}$ exhibits an almost quadratic dependence on temperature, as shown by the grey dashed curve in figure~\ref{Io}. Such a residual temperature dependence of $I_{0}^{CT}$ was theoretically predicted by Hill~\cite{Hill1971b}.

The parameter $T_{0}$ slightly increases with temperature; nevertheless, it has a relative small variation, less than a factor five in a range of 320\,K. The average value of this parameter is $T_{0}^{avg}\approx 2\times10^{6}$\,K. From $T_{0}^{avg}$ one can estimate the density of traps involved in the VRH conduction process by means of the formula
\begin{align}
\label{eq:NtVRH}
N_{T}^{VRH}\approx \frac{36U}{k_{B}T_{0}^{avg}a^{3}}
\end{align}
Assuming $a=2.8$\,nm and $U=135$\,meV we obtained an average density of traps involved in the VRH conduction of about $1\times10^{18}$\,cm$^{-3}$. This value is approximately 20 times higher than the one estimated from DLTS measurements for the deep states $T_{1}$, probably because the VRH involves more traps distributed over a wide range of energies in the band gap. Unfortunately, the lack of experimentally determined values for the parameters $a$ and $U$ casts some doubt on the validity of this estimate. In particular, because of the strong (cubic) dependence of $N_{T}$ on $a$, the value of the density of traps can be adjusted over three orders of magnitude by choosing different values of $a$ (which is unknown) in the range 1$\--$10\,nm. 

The values of the effective electron mass ($m_{e}^{*}$) used to control the slope of the current emitted by CT states vary very little with temperature. The employed values of $m_{e}^{*}$ agree very well with the ones obtained from theoretical~\cite{Fritsch2003} ($m_{e}^{*}=0.13\--0.2\,m_{e}$) and experimental~\cite{Witowski1999} ($m_{e}^{*}=0.22\,m_{e}$ at T=300\,K) studies for GaN. 

\begin{figure}[t]
\includegraphics*[height=6.5cm]{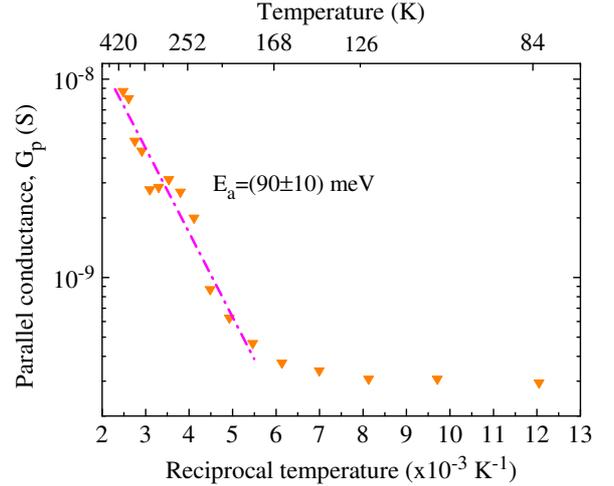} 
\caption{(Color online) Arrhenius plot of the electrical conductance ($G_{p}$) obtained in the low bias regime ($|V|\lesssim 3$\,V) from the empirical parameter $R_{p}$. The magenta dotted-dashed line is a linear interpolation of the data at high temperatures.}
\label{Rp}
\end{figure}

The leakage current in the low bias regime ($|V|\lesssim 3$\,V) is described by the parallel resistance $R_{p}$; for sake of completeness, we discuss also the temperature dependence of this empirical parameter. To this end, we make use of the electrical conductance, defined as the inverse of the parallel resistance: $G_{p}=1/R_{p}$. Figure~\ref{Rp} shows the Arrhenius plot of $G_{p}$. At low temperatures the electrical conductance varies very slowly, whereas for $T\gtrsim 180$\,K a much faster increase of this parameter is observed. This behaviour would suggest the coexistence of different conduction mechanisms also at low biases, one of which is activated at high temperatures. From a linear interpolation of the data in the Arrhenius plot (see magenta dotted-dashed line), we have estimated a thermal activation energy ($E_{a}$) of the conduction mechanism dominant at high temperatures equal to $(90\pm10)$\,meV. However, to clarify the physical origin of this activation energy further studies are required.

In conclusion, we point out that the values of all the physical parameters employed in our model are either comparable to the ones reported in literature or compatible with the expected theoretical behaviour. In addition, they provide a rather good description of the experimental measurements. Therefore, it seems fair to conclude that the developed model provides a reasonable and compelling description of the physical origin of the reverse leakage current in NW-LEDs.

\section{Conclusion}

The transport mechanisms in reversely biased NW-LEDs based on (In,Ga)N/GaN heterostructures have been carefully investigated by means of DLTS and temperature-dependent I-V measurements. The DLTS measurements have revealed the presence of two main deep states at energies of $570\pm20$ and ($840\pm30$)\,meV below the CBM in the active region of NW-LEDs. These defects exhibit a relatively high concentration in the range of $5\times10^{16}$\,cm$^{-3}$. The comparison of the signatures of the observed electron traps with the ones reported in literature has indicated that the deep states might originate from nitrogen antisite defects ($N_{Ga}$), and perhaps also from chains of boundary dislocations (BDs) at the edge of coalescing NWs and dangling bonds present on the free sidewalls. It was recently demonstrated that the latter issue can be significantly mitigated by an adequate surface passivation~\cite{Nguyen2013,Kishino2014a}. The other points are more challenging to overcome. For example, it is well known that III/V flux ratio and crystal polarity play an important role in the formation of point defects~\cite{Hierro2002,Tuomisto2005,Peta2013,Fernandez-Garrido2015}. In particular, a high N flux could promote the formation of Ga vacancies and nitrogen antisite defects. We could thus speculate that lower N fluxes, possibly also characterized by a smaller amount of ions, could reduce the density of native defects in NWs, but it has to be kept in mind that excess N is needed for NW growth. A similar result could also be achieved by growing the NWs at unusually high temperatures~\cite{Zettler2015}. In addition, an improved alignment or reduced number density of the NWs could decrease their coalescence degree and consequently also the formation of the related boundary dislocations. 

The information about deep levels obtained via DLTS measurements has been used to develop a comprehensible physical model able to describe quantitatively the peculiar temperature dependence of the I-V characteristics of reversely biased NW-LEDs. In our model we have combined in a single mathematical expression different types of conduction mechanisms associated to various types of band gap states. In particular, we have found that at moderate and high reverse biases (from -3 to -10\,V), the current conduction can be described by the interaction of two competing mechanisms, namely VRH and emission from CT states described by means of PhaT and the PF effect. These results provide a deeper insight of the charge transport mechanisms present in GaN-based NW-LEDs and suggest a possible way to improve the performance of such devices. Furthermore, the assumptions made in the model as well as the considered transport mechanisms are not bound to the morphology of the sample, thus they should remain valid also for planar devices based on III-N heterostructures. Consequently, the developed model can be in principle applied, adjusting the relevant parameters, also to conventional planar LEDs.

\section{Appendix}\label{sec:app}

In order to determine the optimum value of the fitting parameters, the \textquotedblleft degree of missfit\textquotedblright ($\Xi$) was minimized. This quantity describes the discrepancy between experimental data and calculated curves; it is defined as~\cite{Suzue1996}
\begin{align}
\label{eq:Missfit}
\Xi= \sum_{i=1}^\nu\left|\log\frac{I_{i}^{fit}}{I_{i}^{data}}\right| 
\end{align}
where $\nu$ is the number of measured data points, while $I_{i}^{fit}$ and $I_{i}^{data}$ represent the current at the point $i$ evaluated through equation~\ref{eq:Itot} or measured. The logarithmic function is zero when both contributions are equal. The sum over the entire set of data points provides a good estimate of the quality of the fit.

The reverse bias range can ideally be divided into three regions: low voltages, from 0 to about -3\,V, moderate voltages, roughly between -3 and -6\,V, and high voltages, for biases between -6 and -10\,V. Each bias range is dominated by one conduction process. In particular, the linear contribution defined by $R_P$ describes the low voltage regime, whereas moderate and high bias regions are dominated by VRH and emission from CT states, respectively. For this reason the current in each region can be modelled by means of only one or two parameters. The optimal values of the relevant parameters in each voltage range was found by minimizing $\Xi$. Due to the fact that the parameters are independent, it is rather unlikely to find multiple local minima and, therefore, a single set of parameters can be unequivocally determined. The best fit was deemed to have been achieved when, upon varying any of the parameters by $1\%$ of its value, the calculated current differed from the measured one by a factor smaller than $\pm 10\%$; namely, $\Xi\leq\nu\left|\log(1\pm 0.1)\right|$. Hence, we obtained errors of the fitting parameters smaller than few percent of their values. Note that the sensitivity of the total current to the parameters is rather high; in other words, a small change of their values produces a big variation of the calculated current.

\section{Acknowledgement}

The authors gratefully acknowledge W. Anders for their help with the fabrication of the devices. We are also grateful to J. E. Boschker for a critical reading of the manuscript. This work has been partially supported by the European Commission (project DEEPEN, FP7-NMP-2013-SMALL-7, grant agreement no. 604416). 


\bibliography{bibl_4}

\begin{thebibliography}{58}%
\makeatletter
\providecommand \@ifxundefined [1]{%
 \@ifx{#1\undefined}
}%
\providecommand \@ifnum [1]{%
 \ifnum #1\expandafter \@firstoftwo
 \else \expandafter \@secondoftwo
 \fi
}%
\providecommand \@ifx [1]{%
 \ifx #1\expandafter \@firstoftwo
 \else \expandafter \@secondoftwo
 \fi
}%
\providecommand \natexlab [1]{#1}%
\providecommand \enquote  [1]{``#1''}%
\providecommand \bibnamefont  [1]{#1}%
\providecommand \bibfnamefont [1]{#1}%
\providecommand \citenamefont [1]{#1}%
\providecommand \href@noop [0]{\@secondoftwo}%
\providecommand \href [0]{\begingroup \@sanitize@url \@href}%
\providecommand \@href[1]{\@@startlink{#1}\@@href}%
\providecommand \@@href[1]{\endgroup#1\@@endlink}%
\providecommand \@sanitize@url [0]{\catcode `\\12\catcode `\$12\catcode
  `\&12\catcode `\#12\catcode `\^12\catcode `\_12\catcode `\%12\relax}%
\providecommand \@@startlink[1]{}%
\providecommand \@@endlink[0]{}%
\providecommand \url  [0]{\begingroup\@sanitize@url \@url }%
\providecommand \@url [1]{\endgroup\@href {#1}{\urlprefix }}%
\providecommand \urlprefix  [0]{URL }%
\providecommand \Eprint [0]{\href }%
\providecommand \doibase [0]{http://dx.doi.org/}%
\providecommand \selectlanguage [0]{\@gobble}%
\providecommand \bibinfo  [0]{\@secondoftwo}%
\providecommand \bibfield  [0]{\@secondoftwo}%
\providecommand \translation [1]{[#1]}%
\providecommand \BibitemOpen [0]{}%
\providecommand \bibitemStop [0]{}%
\providecommand \bibitemNoStop [0]{.\EOS\space}%
\providecommand \EOS [0]{\spacefactor3000\relax}%
\providecommand \BibitemShut  [1]{\csname bibitem#1\endcsname}%
\let\auto@bib@innerbib\@empty
\bibitem [{\citenamefont {Li}\ and\ \citenamefont {Waag}(2012)}]{Li2012}%
  \BibitemOpen
  \bibfield  {author} {\bibinfo {author} {\bibfnamefont {S.}~\bibnamefont
  {Li}}\ and\ \bibinfo {author} {\bibfnamefont {A.}~\bibnamefont {Waag}},\
  }\href@noop {} {\bibfield  {journal} {\bibinfo  {journal} {J. Appl. Phys.}\
  }\textbf {\bibinfo {volume} {111}},\ \bibinfo {pages} {071101} (\bibinfo
  {year} {2012})}\BibitemShut {NoStop}%
\bibitem [{\citenamefont {K\"{o}lper}\ \emph {et~al.}(2012)\citenamefont
  {K\"{o}lper}, \citenamefont {Sabathil}, \citenamefont {Mandl}, \citenamefont
  {Strassburg},\ and\ \citenamefont {Witzigmann}}]{Kolper2012}%
  \BibitemOpen
  \bibfield  {author} {\bibinfo {author} {\bibfnamefont {C.}~\bibnamefont
  {K\"{o}lper}}, \bibinfo {author} {\bibfnamefont {M.}~\bibnamefont
  {Sabathil}}, \bibinfo {author} {\bibfnamefont {M.}~\bibnamefont {Mandl}},
  \bibinfo {author} {\bibfnamefont {M.}~\bibnamefont {Strassburg}}, \ and\
  \bibinfo {author} {\bibfnamefont {B.}~\bibnamefont {Witzigmann}},\
  }\href@noop {} {\bibfield  {journal} {\bibinfo  {journal} {J. Lightwave
  Technol.}\ }\textbf {\bibinfo {volume} {30}},\ \bibinfo {pages} {2853}
  (\bibinfo {year} {2012})}\BibitemShut {NoStop}%
\bibitem [{\citenamefont {Kikuchi}\ \emph {et~al.}(2004)\citenamefont
  {Kikuchi}, \citenamefont {Kawai}, \citenamefont {Tada},\ and\ \citenamefont
  {Kishino}}]{Kikuchi2004}%
  \BibitemOpen
  \bibfield  {author} {\bibinfo {author} {\bibfnamefont {A.}~\bibnamefont
  {Kikuchi}}, \bibinfo {author} {\bibfnamefont {M.}~\bibnamefont {Kawai}},
  \bibinfo {author} {\bibfnamefont {M.}~\bibnamefont {Tada}}, \ and\ \bibinfo
  {author} {\bibfnamefont {K.}~\bibnamefont {Kishino}},\ }\href@noop {}
  {\bibfield  {journal} {\bibinfo  {journal} {Jpn. J. Appl. Phys.}\ }\textbf
  {\bibinfo {volume} {43}},\ \bibinfo {pages} {L1524} (\bibinfo {year}
  {2004})}\BibitemShut {NoStop}%
\bibitem [{\citenamefont {Sekiguchi}\ \emph {et~al.}(2008)\citenamefont
  {Sekiguchi}, \citenamefont {Kato}, \citenamefont {Tanaka}, \citenamefont
  {Kikuchi},\ and\ \citenamefont {Kishino}}]{Sekiguchi2008}%
  \BibitemOpen
  \bibfield  {author} {\bibinfo {author} {\bibfnamefont {H.}~\bibnamefont
  {Sekiguchi}}, \bibinfo {author} {\bibfnamefont {K.}~\bibnamefont {Kato}},
  \bibinfo {author} {\bibfnamefont {J.}~\bibnamefont {Tanaka}}, \bibinfo
  {author} {\bibfnamefont {A.}~\bibnamefont {Kikuchi}}, \ and\ \bibinfo
  {author} {\bibfnamefont {K.}~\bibnamefont {Kishino}},\ }\href@noop {}
  {\bibfield  {journal} {\bibinfo  {journal} {Phys. Status Solidi A}\ }\textbf
  {\bibinfo {volume} {205}},\ \bibinfo {pages} {1067} (\bibinfo {year}
  {2008})}\BibitemShut {NoStop}%
\bibitem [{\citenamefont {Lin}\ \emph {et~al.}(2010)\citenamefont {Lin},
  \citenamefont {Lu}, \citenamefont {Chen}, \citenamefont {Lee},\ and\
  \citenamefont {Gwo}}]{Lin2010}%
  \BibitemOpen
  \bibfield  {author} {\bibinfo {author} {\bibfnamefont {H.-W.}\ \bibnamefont
  {Lin}}, \bibinfo {author} {\bibfnamefont {Y.-J.}\ \bibnamefont {Lu}},
  \bibinfo {author} {\bibfnamefont {H.-Y.}\ \bibnamefont {Chen}}, \bibinfo
  {author} {\bibfnamefont {H.-M.}\ \bibnamefont {Lee}}, \ and\ \bibinfo
  {author} {\bibfnamefont {S.}~\bibnamefont {Gwo}},\ }\href@noop {} {\bibfield
  {journal} {\bibinfo  {journal} {Appl. Phys. Lett.}\ }\textbf {\bibinfo
  {volume} {97}},\ \bibinfo {pages} {073101} (\bibinfo {year}
  {2010})}\BibitemShut {NoStop}%
\bibitem [{\citenamefont {Guo}\ \emph {et~al.}(2010)\citenamefont {Guo},
  \citenamefont {Zhang}, \citenamefont {Banerjee},\ and\ \citenamefont
  {Bhattacharya}}]{Guo2010}%
  \BibitemOpen
  \bibfield  {author} {\bibinfo {author} {\bibfnamefont {W.}~\bibnamefont
  {Guo}}, \bibinfo {author} {\bibfnamefont {M.}~\bibnamefont {Zhang}}, \bibinfo
  {author} {\bibfnamefont {A.}~\bibnamefont {Banerjee}}, \ and\ \bibinfo
  {author} {\bibfnamefont {P.}~\bibnamefont {Bhattacharya}},\ }\href@noop {}
  {\bibfield  {journal} {\bibinfo  {journal} {Nano Lett.}\ }\textbf {\bibinfo
  {volume} {10}},\ \bibinfo {pages} {3355} (\bibinfo {year}
  {2010})}\BibitemShut {NoStop}%
\bibitem [{\citenamefont {Bavencove}\ \emph {et~al.}(2010)\citenamefont
  {Bavencove}, \citenamefont {Tourbot}, \citenamefont {Pougeoise},
  \citenamefont {Garcia}, \citenamefont {Gilet}, \citenamefont {Levy},
  \citenamefont {Andr\'{e}}, \citenamefont {Feuillet}, \citenamefont {Gayral},
  \citenamefont {Daudin},\ and\ \citenamefont {Dang}}]{Bavencove2010}%
  \BibitemOpen
  \bibfield  {author} {\bibinfo {author} {\bibfnamefont {A.-L.}\ \bibnamefont
  {Bavencove}}, \bibinfo {author} {\bibfnamefont {G.}~\bibnamefont {Tourbot}},
  \bibinfo {author} {\bibfnamefont {E.}~\bibnamefont {Pougeoise}}, \bibinfo
  {author} {\bibfnamefont {J.}~\bibnamefont {Garcia}}, \bibinfo {author}
  {\bibfnamefont {P.}~\bibnamefont {Gilet}}, \bibinfo {author} {\bibfnamefont
  {F.}~\bibnamefont {Levy}}, \bibinfo {author} {\bibfnamefont {B.}~\bibnamefont
  {Andr\'{e}}}, \bibinfo {author} {\bibfnamefont {G.}~\bibnamefont {Feuillet}},
  \bibinfo {author} {\bibfnamefont {B.}~\bibnamefont {Gayral}}, \bibinfo
  {author} {\bibfnamefont {B.}~\bibnamefont {Daudin}}, \ and\ \bibinfo {author}
  {\bibfnamefont {L.~S.}\ \bibnamefont {Dang}},\ }\href@noop {} {\bibfield
  {journal} {\bibinfo  {journal} {Phys. Status Solidi A}\ }\textbf {\bibinfo
  {volume} {207}},\ \bibinfo {pages} {1425} (\bibinfo {year}
  {2010})}\BibitemShut {NoStop}%
\bibitem [{\citenamefont {Armitage}\ and\ \citenamefont
  {Tsubaki}(2010)}]{Armitage2010}%
  \BibitemOpen
  \bibfield  {author} {\bibinfo {author} {\bibfnamefont {R.}~\bibnamefont
  {Armitage}}\ and\ \bibinfo {author} {\bibfnamefont {K.}~\bibnamefont
  {Tsubaki}},\ }\href@noop {} {\bibfield  {journal} {\bibinfo  {journal}
  {Nanotechnology}\ }\textbf {\bibinfo {volume} {21}},\ \bibinfo {pages}
  {195202} (\bibinfo {year} {2010})}\BibitemShut {NoStop}%
\bibitem [{\citenamefont {Nguyen}\ \emph {et~al.}(2011)\citenamefont {Nguyen},
  \citenamefont {Zhang}, \citenamefont {Cui}, \citenamefont {Han},
  \citenamefont {Fathololoumi}, \citenamefont {Couillard}, \citenamefont
  {Botton},\ and\ \citenamefont {Mi}}]{Nguyen2011}%
  \BibitemOpen
  \bibfield  {author} {\bibinfo {author} {\bibfnamefont {H.~P.~T.}\
  \bibnamefont {Nguyen}}, \bibinfo {author} {\bibfnamefont {S.}~\bibnamefont
  {Zhang}}, \bibinfo {author} {\bibfnamefont {K.}~\bibnamefont {Cui}}, \bibinfo
  {author} {\bibfnamefont {X.}~\bibnamefont {Han}}, \bibinfo {author}
  {\bibfnamefont {S.}~\bibnamefont {Fathololoumi}}, \bibinfo {author}
  {\bibfnamefont {M.}~\bibnamefont {Couillard}}, \bibinfo {author}
  {\bibfnamefont {G.}~\bibnamefont {Botton}}, \ and\ \bibinfo {author}
  {\bibfnamefont {Z.}~\bibnamefont {Mi}},\ }\href@noop {} {\bibfield  {journal}
  {\bibinfo  {journal} {Nano Lett.}\ }\textbf {\bibinfo {volume} {11}},\
  \bibinfo {pages} {1919} (\bibinfo {year} {2011})}\BibitemShut {NoStop}%
\bibitem [{\citenamefont {Ra}\ \emph {et~al.}(2013)\citenamefont {Ra},
  \citenamefont {Navamathavan}, \citenamefont {Park},\ and\ \citenamefont
  {Lee}}]{Ra2013}%
  \BibitemOpen
  \bibfield  {author} {\bibinfo {author} {\bibfnamefont {Y.-H.}\ \bibnamefont
  {Ra}}, \bibinfo {author} {\bibfnamefont {R.}~\bibnamefont {Navamathavan}},
  \bibinfo {author} {\bibfnamefont {J.-H.}\ \bibnamefont {Park}}, \ and\
  \bibinfo {author} {\bibfnamefont {C.-R.}\ \bibnamefont {Lee}},\ }\href@noop
  {} {\bibfield  {journal} {\bibinfo  {journal} {ACS applied materials \&
  interfaces}\ }\textbf {\bibinfo {volume} {5}},\ \bibinfo {pages} {2111}
  (\bibinfo {year} {2013})}\BibitemShut {NoStop}%
\bibitem [{\citenamefont {Musolino}\ \emph {et~al.}(2014)\citenamefont
  {Musolino}, \citenamefont {Tahraoui}, \citenamefont {Limbach}, \citenamefont
  {L\"{a}hnemann}, \citenamefont {Jahn}, \citenamefont {Brandt}, \citenamefont
  {Geelhaar},\ and\ \citenamefont {Riechert}}]{Musolino2014}%
  \BibitemOpen
  \bibfield  {author} {\bibinfo {author} {\bibfnamefont {M.}~\bibnamefont
  {Musolino}}, \bibinfo {author} {\bibfnamefont {A.}~\bibnamefont {Tahraoui}},
  \bibinfo {author} {\bibfnamefont {F.}~\bibnamefont {Limbach}}, \bibinfo
  {author} {\bibfnamefont {J.}~\bibnamefont {L\"{a}hnemann}}, \bibinfo {author}
  {\bibfnamefont {U.}~\bibnamefont {Jahn}}, \bibinfo {author} {\bibfnamefont
  {O.}~\bibnamefont {Brandt}}, \bibinfo {author} {\bibfnamefont
  {L.}~\bibnamefont {Geelhaar}}, \ and\ \bibinfo {author} {\bibfnamefont
  {H.}~\bibnamefont {Riechert}},\ }\href@noop {} {\bibfield  {journal}
  {\bibinfo  {journal} {Appl. Phys. Lett.}\ }\textbf {\bibinfo {volume}
  {105}},\ \bibinfo {pages} {083505} (\bibinfo {year} {2014})}\BibitemShut
  {NoStop}%
\bibitem [{\citenamefont {Zhang}\ \emph {et~al.}(2014)\citenamefont {Zhang},
  \citenamefont {Connie}, \citenamefont {Laleyan}, \citenamefont {Nguyen},
  \citenamefont {Wang}, \citenamefont {Song}, \citenamefont {Shih},\ and\
  \citenamefont {Mi}}]{Zhang2014e}%
  \BibitemOpen
  \bibfield  {author} {\bibinfo {author} {\bibfnamefont {S.}~\bibnamefont
  {Zhang}}, \bibinfo {author} {\bibfnamefont {A.~T.}\ \bibnamefont {Connie}},
  \bibinfo {author} {\bibfnamefont {D.~A.}\ \bibnamefont {Laleyan}}, \bibinfo
  {author} {\bibfnamefont {H.~P.~T.}\ \bibnamefont {Nguyen}}, \bibinfo {author}
  {\bibfnamefont {Q.}~\bibnamefont {Wang}}, \bibinfo {author} {\bibfnamefont
  {J.}~\bibnamefont {Song}}, \bibinfo {author} {\bibfnamefont {I.}~\bibnamefont
  {Shih}}, \ and\ \bibinfo {author} {\bibfnamefont {Z.}~\bibnamefont {Mi}},\
  }\href@noop {} {\bibfield  {journal} {\bibinfo  {journal} {IEEE J. Quantum
  Electron.}\ }\textbf {\bibinfo {volume} {50}},\ \bibinfo {pages} {483}
  (\bibinfo {year} {2014})}\BibitemShut {NoStop}%
\bibitem [{\citenamefont {Jahangir}\ \emph {et~al.}(2014)\citenamefont
  {Jahangir}, \citenamefont {Schimpke}, \citenamefont {Strassburg},
  \citenamefont {Grossklaus}, \citenamefont {Millunchick},\ and\ \citenamefont
  {Bhattacharya}}]{Jahangir2014}%
  \BibitemOpen
  \bibfield  {author} {\bibinfo {author} {\bibfnamefont {S.}~\bibnamefont
  {Jahangir}}, \bibinfo {author} {\bibfnamefont {T.}~\bibnamefont {Schimpke}},
  \bibinfo {author} {\bibfnamefont {M.}~\bibnamefont {Strassburg}}, \bibinfo
  {author} {\bibfnamefont {K.~A.}\ \bibnamefont {Grossklaus}}, \bibinfo
  {author} {\bibfnamefont {J.~M.}\ \bibnamefont {Millunchick}}, \ and\ \bibinfo
  {author} {\bibfnamefont {P.}~\bibnamefont {Bhattacharya}},\ }\href@noop {}
  {\bibfield  {journal} {\bibinfo  {journal} {IEEE J. Quantum Electron.}\
  }\textbf {\bibinfo {volume} {50}},\ \bibinfo {pages} {530} (\bibinfo {year}
  {2014})}\BibitemShut {NoStop}%
\bibitem [{\citenamefont {Musolino}\ \emph
  {et~al.}(2015{\natexlab{a}})\citenamefont {Musolino}, \citenamefont
  {Meneghini}, \citenamefont {Scarparo}, \citenamefont {{De Santi}},
  \citenamefont {Tahraoui}, \citenamefont {Geelhaar}, \citenamefont {Zanoni},\
  and\ \citenamefont {Riechert}}]{Musolino2015b}%
  \BibitemOpen
  \bibfield  {author} {\bibinfo {author} {\bibfnamefont {M.}~\bibnamefont
  {Musolino}}, \bibinfo {author} {\bibfnamefont {M.}~\bibnamefont {Meneghini}},
  \bibinfo {author} {\bibfnamefont {L.}~\bibnamefont {Scarparo}}, \bibinfo
  {author} {\bibfnamefont {C.}~\bibnamefont {{De Santi}}}, \bibinfo {author}
  {\bibfnamefont {A.}~\bibnamefont {Tahraoui}}, \bibinfo {author}
  {\bibfnamefont {L.}~\bibnamefont {Geelhaar}}, \bibinfo {author}
  {\bibfnamefont {E.}~\bibnamefont {Zanoni}}, \ and\ \bibinfo {author}
  {\bibfnamefont {H.}~\bibnamefont {Riechert}},\ }\href@noop {} {\bibfield
  {journal} {\bibinfo  {journal} {Proc. SPIE}\ }\textbf {\bibinfo {volume}
  {9363}},\ \bibinfo {pages} {936325} (\bibinfo {year}
  {2015}{\natexlab{a}})}\BibitemShut {NoStop}%
\bibitem [{\citenamefont {Casey}\ \emph {et~al.}(1996)\citenamefont {Casey},
  \citenamefont {Muth}, \citenamefont {Krishnankutty},\ and\ \citenamefont
  {Zavada}}]{Casey1996}%
  \BibitemOpen
  \bibfield  {author} {\bibinfo {author} {\bibfnamefont {H.~C.}\ \bibnamefont
  {Casey}}, \bibinfo {author} {\bibfnamefont {J.}~\bibnamefont {Muth}},
  \bibinfo {author} {\bibfnamefont {S.}~\bibnamefont {Krishnankutty}}, \ and\
  \bibinfo {author} {\bibfnamefont {J.~M.}\ \bibnamefont {Zavada}},\
  }\href@noop {} {\bibfield  {journal} {\bibinfo  {journal} {Appl. Phys. Lett}\
  }\textbf {\bibinfo {volume} {68}},\ \bibinfo {pages} {2867} (\bibinfo {year}
  {1996})}\BibitemShut {NoStop}%
\bibitem [{\citenamefont {Yan}\ \emph {et~al.}(2010)\citenamefont {Yan},
  \citenamefont {Lu}, \citenamefont {Chen}, \citenamefont {Zhang},\ and\
  \citenamefont {Zheng}}]{Yan2010}%
  \BibitemOpen
  \bibfield  {author} {\bibinfo {author} {\bibfnamefont {D.}~\bibnamefont
  {Yan}}, \bibinfo {author} {\bibfnamefont {H.}~\bibnamefont {Lu}}, \bibinfo
  {author} {\bibfnamefont {D.}~\bibnamefont {Chen}}, \bibinfo {author}
  {\bibfnamefont {R.}~\bibnamefont {Zhang}}, \ and\ \bibinfo {author}
  {\bibfnamefont {Y.}~\bibnamefont {Zheng}},\ }\href@noop {} {\bibfield
  {journal} {\bibinfo  {journal} {Appl Phys. Lett.}\ }\textbf {\bibinfo
  {volume} {96}},\ \bibinfo {pages} {083504} (\bibinfo {year}
  {2010})}\BibitemShut {NoStop}%
\bibitem [{\citenamefont {Shan}\ \emph {et~al.}(2011)\citenamefont {Shan},
  \citenamefont {Meyaard}, \citenamefont {Dai}, \citenamefont {Cho},
  \citenamefont {Schubert}, \citenamefont {Son},\ and\ \citenamefont
  {Sone}}]{Shan2011}%
  \BibitemOpen
  \bibfield  {author} {\bibinfo {author} {\bibfnamefont {Q.}~\bibnamefont
  {Shan}}, \bibinfo {author} {\bibfnamefont {D.~S.}\ \bibnamefont {Meyaard}},
  \bibinfo {author} {\bibfnamefont {Q.}~\bibnamefont {Dai}}, \bibinfo {author}
  {\bibfnamefont {J.}~\bibnamefont {Cho}}, \bibinfo {author} {\bibfnamefont
  {F.~E.}\ \bibnamefont {Schubert}}, \bibinfo {author} {\bibfnamefont {K.~J.}\
  \bibnamefont {Son}}, \ and\ \bibinfo {author} {\bibfnamefont
  {C.}~\bibnamefont {Sone}},\ }\href@noop {} {\bibfield  {journal} {\bibinfo
  {journal} {Appl. Phys. Lett.}\ }\textbf {\bibinfo {volume} {99}},\ \bibinfo
  {pages} {253506} (\bibinfo {year} {2011})}\BibitemShut {NoStop}%
\bibitem [{\citenamefont {Meneghini}\ \emph {et~al.}(2014)\citenamefont
  {Meneghini}, \citenamefont {la~Grassa}, \citenamefont {Vaccari},
  \citenamefont {Galler}, \citenamefont {Zeisel}, \citenamefont {Drechsel},
  \citenamefont {Hahn}, \citenamefont {Meneghesso},\ and\ \citenamefont
  {Zanoni}}]{Meneghini2014}%
  \BibitemOpen
  \bibfield  {author} {\bibinfo {author} {\bibfnamefont {M.}~\bibnamefont
  {Meneghini}}, \bibinfo {author} {\bibfnamefont {M.}~\bibnamefont
  {la~Grassa}}, \bibinfo {author} {\bibfnamefont {S.}~\bibnamefont {Vaccari}},
  \bibinfo {author} {\bibfnamefont {B.}~\bibnamefont {Galler}}, \bibinfo
  {author} {\bibfnamefont {R.}~\bibnamefont {Zeisel}}, \bibinfo {author}
  {\bibfnamefont {P.}~\bibnamefont {Drechsel}}, \bibinfo {author}
  {\bibfnamefont {B.}~\bibnamefont {Hahn}}, \bibinfo {author} {\bibfnamefont
  {G.}~\bibnamefont {Meneghesso}}, \ and\ \bibinfo {author} {\bibfnamefont
  {E.}~\bibnamefont {Zanoni}},\ }\href@noop {} {\bibfield  {journal} {\bibinfo
  {journal} {Appl. Phys. Lett.}\ }\textbf {\bibinfo {volume} {104}},\ \bibinfo
  {pages} {113505} (\bibinfo {year} {2014})}\BibitemShut {NoStop}%
\bibitem [{\citenamefont {{Auf der Maur}}\ \emph {et~al.}(2014)\citenamefont
  {{Auf der Maur}}, \citenamefont {Galler}, \citenamefont {Pietzonka},
  \citenamefont {Strassburg}, \citenamefont {Lugauer},\ and\ \citenamefont {{Di
  Carlo}}}]{AufderMaur2014}%
  \BibitemOpen
  \bibfield  {author} {\bibinfo {author} {\bibfnamefont {M.}~\bibnamefont {{Auf
  der Maur}}}, \bibinfo {author} {\bibfnamefont {B.}~\bibnamefont {Galler}},
  \bibinfo {author} {\bibfnamefont {I.}~\bibnamefont {Pietzonka}}, \bibinfo
  {author} {\bibfnamefont {M.}~\bibnamefont {Strassburg}}, \bibinfo {author}
  {\bibfnamefont {H.}~\bibnamefont {Lugauer}}, \ and\ \bibinfo {author}
  {\bibfnamefont {A.}~\bibnamefont {{Di Carlo}}},\ }\href@noop {} {\bibfield
  {journal} {\bibinfo  {journal} {Appl. Phys. Lett.}\ }\textbf {\bibinfo
  {volume} {105}},\ \bibinfo {pages} {133504} (\bibinfo {year}
  {2014})}\BibitemShut {NoStop}%
\bibitem [{\citenamefont {Lee}\ \emph {et~al.}(2011)\citenamefont {Lee},
  \citenamefont {Lee}, \citenamefont {Chen}, \citenamefont {Lu},\ and\
  \citenamefont {Kuo}}]{Lee2011a}%
  \BibitemOpen
  \bibfield  {author} {\bibinfo {author} {\bibfnamefont {Y.-J.}\ \bibnamefont
  {Lee}}, \bibinfo {author} {\bibfnamefont {C.-J.}\ \bibnamefont {Lee}},
  \bibinfo {author} {\bibfnamefont {C.-H.}\ \bibnamefont {Chen}}, \bibinfo
  {author} {\bibfnamefont {T.-C.}\ \bibnamefont {Lu}}, \ and\ \bibinfo {author}
  {\bibfnamefont {H.-C.}\ \bibnamefont {Kuo}},\ }\href@noop {} {\bibfield
  {journal} {\bibinfo  {journal} {IEEE J. Sel. Topics Quantum Electron.}\
  }\textbf {\bibinfo {volume} {17}},\ \bibinfo {pages} {985} (\bibinfo {year}
  {2011})}\BibitemShut {NoStop}%
\bibitem [{\citenamefont {Kishino}, \citenamefont {Kamimura},\ and\
  \citenamefont {Kamiyama}(2012)}]{Kishino2012}%
  \BibitemOpen
  \bibfield  {author} {\bibinfo {author} {\bibfnamefont {K.}~\bibnamefont
  {Kishino}}, \bibinfo {author} {\bibfnamefont {J.}~\bibnamefont {Kamimura}}, \
  and\ \bibinfo {author} {\bibfnamefont {K.}~\bibnamefont {Kamiyama}},\
  }\href@noop {} {\bibfield  {journal} {\bibinfo  {journal} {Appl. Phys.
  Express}\ }\textbf {\bibinfo {volume} {5}},\ \bibinfo {pages} {2} (\bibinfo
  {year} {2012})}\BibitemShut {NoStop}%
\bibitem [{\citenamefont {Tchernycheva}\ \emph {et~al.}(2014)\citenamefont
  {Tchernycheva}, \citenamefont {Lavenus}, \citenamefont {Zhang}, \citenamefont
  {Babichev}, \citenamefont {Jacopin}, \citenamefont {Shahmohammadi},
  \citenamefont {Julien}, \citenamefont {Ciechonski}, \citenamefont {Vescovi},\
  and\ \citenamefont {Kryliouk}}]{Tchernycheva2014}%
  \BibitemOpen
  \bibfield  {author} {\bibinfo {author} {\bibfnamefont {M.}~\bibnamefont
  {Tchernycheva}}, \bibinfo {author} {\bibfnamefont {P.}~\bibnamefont
  {Lavenus}}, \bibinfo {author} {\bibfnamefont {H.}~\bibnamefont {Zhang}},
  \bibinfo {author} {\bibfnamefont {A.~V.}\ \bibnamefont {Babichev}}, \bibinfo
  {author} {\bibfnamefont {G.}~\bibnamefont {Jacopin}}, \bibinfo {author}
  {\bibfnamefont {M.}~\bibnamefont {Shahmohammadi}}, \bibinfo {author}
  {\bibfnamefont {F.~H.}\ \bibnamefont {Julien}}, \bibinfo {author}
  {\bibfnamefont {R.}~\bibnamefont {Ciechonski}}, \bibinfo {author}
  {\bibfnamefont {G.}~\bibnamefont {Vescovi}}, \ and\ \bibinfo {author}
  {\bibfnamefont {O.}~\bibnamefont {Kryliouk}},\ }\href@noop {} {\bibfield
  {journal} {\bibinfo  {journal} {Nano Lett.}\ }\textbf {\bibinfo {volume}
  {14}},\ \bibinfo {pages} {2456} (\bibinfo {year} {2014})}\BibitemShut
  {NoStop}%
\bibitem [{\citenamefont {Musolino}\ \emph
  {et~al.}(2015{\natexlab{b}})\citenamefont {Musolino}, \citenamefont
  {Tahraoui}, \citenamefont {Fern\'{a}ndez-Garrido}, \citenamefont {Brandt},
  \citenamefont {Trampert}, \citenamefont {Geelhaar},\ and\ \citenamefont
  {Riechert}}]{Musolino2015a}%
  \BibitemOpen
  \bibfield  {author} {\bibinfo {author} {\bibfnamefont {M.}~\bibnamefont
  {Musolino}}, \bibinfo {author} {\bibfnamefont {A.}~\bibnamefont {Tahraoui}},
  \bibinfo {author} {\bibfnamefont {S.}~\bibnamefont {Fern\'{a}ndez-Garrido}},
  \bibinfo {author} {\bibfnamefont {O.}~\bibnamefont {Brandt}}, \bibinfo
  {author} {\bibfnamefont {A.}~\bibnamefont {Trampert}}, \bibinfo {author}
  {\bibfnamefont {L.}~\bibnamefont {Geelhaar}}, \ and\ \bibinfo {author}
  {\bibfnamefont {H.}~\bibnamefont {Riechert}},\ }\href@noop {} {\bibfield
  {journal} {\bibinfo  {journal} {Nanotechnology}\ }\textbf {\bibinfo {volume}
  {26}},\ \bibinfo {pages} {085605} (\bibinfo {year}
  {2015}{\natexlab{b}})}\BibitemShut {NoStop}%
\bibitem [{\citenamefont {Cao}\ \emph {et~al.}(2003)\citenamefont {Cao},
  \citenamefont {Sandvik}, \citenamefont {LeBoeuf},\ and\ \citenamefont
  {Arthur}}]{Cao2003c}%
  \BibitemOpen
  \bibfield  {author} {\bibinfo {author} {\bibfnamefont {X.~A.}\ \bibnamefont
  {Cao}}, \bibinfo {author} {\bibfnamefont {P.~M.}\ \bibnamefont {Sandvik}},
  \bibinfo {author} {\bibfnamefont {S.~F.}\ \bibnamefont {LeBoeuf}}, \ and\
  \bibinfo {author} {\bibfnamefont {S.~D.}\ \bibnamefont {Arthur}},\
  }\href@noop {} {\bibfield  {journal} {\bibinfo  {journal} {Microelectron.
  Reliab.}\ }\textbf {\bibinfo {volume} {43}},\ \bibinfo {pages} {1987}
  (\bibinfo {year} {2003})}\BibitemShut {NoStop}%
\bibitem [{Cor()}]{Correction}%
  \BibitemOpen
  \href@noop {} {\bibinfo  {journal} {The DLTS measurements have already been
  published in Ref.~14. Meanwhile, it has come to our attention that the
  corresponding raw data were not normalized properly. For the results
  presented here, the necessary correction has been carried out}\ }\BibitemShut
  {NoStop}%
\bibitem [{\citenamefont {Hierro}\ \emph {et~al.}(2001)\citenamefont {Hierro},
  \citenamefont {Arehart}, \citenamefont {Heying}, \citenamefont {Hansen},
  \citenamefont {Speck}, \citenamefont {Mishra}, \citenamefont {DenBaars},\
  and\ \citenamefont {Ringel}}]{Hierro2001}%
  \BibitemOpen
\bibfield  {journal} {  }\bibfield  {author} {\bibinfo {author} {\bibfnamefont
  {A.}~\bibnamefont {Hierro}}, \bibinfo {author} {\bibfnamefont
  {A.}~\bibnamefont {Arehart}}, \bibinfo {author} {\bibfnamefont
  {B.}~\bibnamefont {Heying}}, \bibinfo {author} {\bibfnamefont
  {M.}~\bibnamefont {Hansen}}, \bibinfo {author} {\bibfnamefont
  {J.}~\bibnamefont {Speck}}, \bibinfo {author} {\bibfnamefont
  {U.}~\bibnamefont {Mishra}}, \bibinfo {author} {\bibfnamefont
  {S.}~\bibnamefont {DenBaars}}, \ and\ \bibinfo {author} {\bibfnamefont
  {S.}~\bibnamefont {Ringel}},\ }\href@noop {} {\bibfield  {journal} {\bibinfo
  {journal} {Phys. Status Solidi B}\ }\textbf {\bibinfo {volume} {228}},\
  \bibinfo {pages} {309} (\bibinfo {year} {2001})}\BibitemShut {NoStop}%
\bibitem [{\citenamefont {Osaka}\ \emph {et~al.}(2005)\citenamefont {Osaka},
  \citenamefont {Ohno}, \citenamefont {Kishimoto}, \citenamefont {Maezawa},\
  and\ \citenamefont {Mizutani}}]{Osaka2005}%
  \BibitemOpen
  \bibfield  {author} {\bibinfo {author} {\bibfnamefont {J.}~\bibnamefont
  {Osaka}}, \bibinfo {author} {\bibfnamefont {Y.}~\bibnamefont {Ohno}},
  \bibinfo {author} {\bibfnamefont {S.}~\bibnamefont {Kishimoto}}, \bibinfo
  {author} {\bibfnamefont {K.}~\bibnamefont {Maezawa}}, \ and\ \bibinfo
  {author} {\bibfnamefont {T.}~\bibnamefont {Mizutani}},\ }\href@noop {}
  {\bibfield  {journal} {\bibinfo  {journal} {Appl. Phys. Lett.}\ }\textbf
  {\bibinfo {volume} {87}},\ \bibinfo {pages} {222112} (\bibinfo {year}
  {2005})}\BibitemShut {NoStop}%
\bibitem [{\citenamefont {Hacke}\ \emph {et~al.}(1994)\citenamefont {Hacke},
  \citenamefont {Detchprohm}, \citenamefont {Hiramatsu}, \citenamefont
  {Sawaki}, \citenamefont {Tadatomo},\ and\ \citenamefont
  {Miyake}}]{Hacke1994}%
  \BibitemOpen
  \bibfield  {author} {\bibinfo {author} {\bibfnamefont {P.}~\bibnamefont
  {Hacke}}, \bibinfo {author} {\bibfnamefont {T.}~\bibnamefont {Detchprohm}},
  \bibinfo {author} {\bibfnamefont {K.}~\bibnamefont {Hiramatsu}}, \bibinfo
  {author} {\bibfnamefont {N.}~\bibnamefont {Sawaki}}, \bibinfo {author}
  {\bibfnamefont {K.}~\bibnamefont {Tadatomo}}, \ and\ \bibinfo {author}
  {\bibfnamefont {K.}~\bibnamefont {Miyake}},\ }\href@noop {} {\bibfield
  {journal} {\bibinfo  {journal} {J. Appl. Phys.}\ }\textbf {\bibinfo {volume}
  {76}},\ \bibinfo {pages} {304} (\bibinfo {year} {1994})}\BibitemShut
  {NoStop}%
\bibitem [{\citenamefont {Haase}\ \emph {et~al.}(1996)\citenamefont {Haase},
  \citenamefont {Schmid}, \citenamefont {K\"urner}, \citenamefont {D\"ornen},
  \citenamefont {H\"arle}, \citenamefont {Scholz}, \citenamefont {Burkard},\
  and\ \citenamefont {Schweizer}}]{Haase1996}%
  \BibitemOpen
  \bibfield  {author} {\bibinfo {author} {\bibfnamefont {D.}~\bibnamefont
  {Haase}}, \bibinfo {author} {\bibfnamefont {M.}~\bibnamefont {Schmid}},
  \bibinfo {author} {\bibfnamefont {W.}~\bibnamefont {K\"urner}}, \bibinfo
  {author} {\bibfnamefont {A.}~\bibnamefont {D\"ornen}}, \bibinfo {author}
  {\bibfnamefont {V.}~\bibnamefont {H\"arle}}, \bibinfo {author} {\bibfnamefont
  {F.}~\bibnamefont {Scholz}}, \bibinfo {author} {\bibfnamefont
  {M.}~\bibnamefont {Burkard}}, \ and\ \bibinfo {author} {\bibfnamefont
  {H.}~\bibnamefont {Schweizer}},\ }\href@noop {} {\bibfield  {journal}
  {\bibinfo  {journal} {Appl. Phys. Lett.}\ }\textbf {\bibinfo {volume} {69}},\
  \bibinfo {pages} {2525} (\bibinfo {year} {1996})}\BibitemShut {NoStop}%
\bibitem [{\citenamefont {Wang}\ \emph {et~al.}(1998)\citenamefont {Wang},
  \citenamefont {Yu}, \citenamefont {Lau}, \citenamefont {Yu}, \citenamefont
  {Kim}, \citenamefont {Botchkarev},\ and\ \citenamefont
  {Morko\c{c}}}]{Wang1998}%
  \BibitemOpen
  \bibfield  {author} {\bibinfo {author} {\bibfnamefont {C.~D.}\ \bibnamefont
  {Wang}}, \bibinfo {author} {\bibfnamefont {L.~S.}\ \bibnamefont {Yu}},
  \bibinfo {author} {\bibfnamefont {S.~S.}\ \bibnamefont {Lau}}, \bibinfo
  {author} {\bibfnamefont {E.~T.}\ \bibnamefont {Yu}}, \bibinfo {author}
  {\bibfnamefont {W.}~\bibnamefont {Kim}}, \bibinfo {author} {\bibfnamefont
  {A.~E.}\ \bibnamefont {Botchkarev}}, \ and\ \bibinfo {author} {\bibfnamefont
  {H.}~\bibnamefont {Morko\c{c}}},\ }\href@noop {} {\bibfield  {journal}
  {\bibinfo  {journal} {Appl. Phys. Lett.}\ }\textbf {\bibinfo {volume} {72}},\
  \bibinfo {pages} {1211} (\bibinfo {year} {1998})}\BibitemShut {NoStop}%
\bibitem [{\citenamefont {Umana-Membreno}\ \emph {et~al.}(2007)\citenamefont
  {Umana-Membreno}, \citenamefont {Parish}, \citenamefont {Fichtenbaum},
  \citenamefont {Keller}, \citenamefont {Mishra},\ and\ \citenamefont
  {Nener}}]{Umana-Membreno2007}%
  \BibitemOpen
  \bibfield  {author} {\bibinfo {author} {\bibfnamefont {G.~A.}\ \bibnamefont
  {Umana-Membreno}}, \bibinfo {author} {\bibfnamefont {G.}~\bibnamefont
  {Parish}}, \bibinfo {author} {\bibfnamefont {N.}~\bibnamefont {Fichtenbaum}},
  \bibinfo {author} {\bibfnamefont {S.}~\bibnamefont {Keller}}, \bibinfo
  {author} {\bibfnamefont {U.~K.}\ \bibnamefont {Mishra}}, \ and\ \bibinfo
  {author} {\bibfnamefont {B.~D.}\ \bibnamefont {Nener}},\ }\href@noop {}
  {\bibfield  {journal} {\bibinfo  {journal} {J. Electron. Mater.}\ }\textbf
  {\bibinfo {volume} {37}},\ \bibinfo {pages} {569} (\bibinfo {year}
  {2007})}\BibitemShut {NoStop}%
\bibitem [{\citenamefont {Kindl}\ \emph {et~al.}(2009)\citenamefont {Kindl},
  \citenamefont {Hub\'ik}, \citenamefont {Kri\v{s}tofik}, \citenamefont
  {Marev{s}}, \citenamefont {V\'yborn\'y}, \citenamefont {Leys},\ and\
  \citenamefont {Boeykens}}]{Kindl2009}%
  \BibitemOpen
  \bibfield  {author} {\bibinfo {author} {\bibfnamefont {D.}~\bibnamefont
  {Kindl}}, \bibinfo {author} {\bibfnamefont {P.}~\bibnamefont {Hub\'ik}},
  \bibinfo {author} {\bibfnamefont {J.}~\bibnamefont {Kri\v{s}tofik}}, \bibinfo
  {author} {\bibfnamefont {J.~J.}\ \bibnamefont {Marev{s}}}, \bibinfo {author}
  {\bibfnamefont {Z.}~\bibnamefont {V\'yborn\'y}}, \bibinfo {author}
  {\bibfnamefont {M.~R.}\ \bibnamefont {Leys}}, \ and\ \bibinfo {author}
  {\bibfnamefont {S.}~\bibnamefont {Boeykens}},\ }\href@noop {} {\bibfield
  {journal} {\bibinfo  {journal} {J. Appl. Phys.}\ }\textbf {\bibinfo {volume}
  {105}},\ \bibinfo {pages} {093706} (\bibinfo {year} {2009})}\BibitemShut
  {NoStop}%
\bibitem [{\citenamefont {Peta}\ \emph {et~al.}(2013)\citenamefont {Peta},
  \citenamefont {Lee}, \citenamefont {Moon-Deock}, \citenamefont {Oh},
  \citenamefont {Kim},\ and\ \citenamefont {Kim}}]{Peta2013}%
  \BibitemOpen
  \bibfield  {author} {\bibinfo {author} {\bibfnamefont {K.~R.}\ \bibnamefont
  {Peta}}, \bibinfo {author} {\bibfnamefont {S.-T.}\ \bibnamefont {Lee}},
  \bibinfo {author} {\bibfnamefont {K.}~\bibnamefont {Moon-Deock}}, \bibinfo
  {author} {\bibfnamefont {J.-E.}\ \bibnamefont {Oh}}, \bibinfo {author}
  {\bibfnamefont {S.-G.}\ \bibnamefont {Kim}}, \ and\ \bibinfo {author}
  {\bibfnamefont {T.-G.}\ \bibnamefont {Kim}},\ }\href@noop {} {\bibfield
  {journal} {\bibinfo  {journal} {J. Cryst. Growth}\ }\textbf {\bibinfo
  {volume} {378}},\ \bibinfo {pages} {299} (\bibinfo {year}
  {2013})}\BibitemShut {NoStop}%
\bibitem [{\citenamefont {Consonni}\ \emph {et~al.}(2009)\citenamefont
  {Consonni}, \citenamefont {Knelangen}, \citenamefont {Jahn}, \citenamefont
  {Trampert}, \citenamefont {Geelhaar},\ and\ \citenamefont
  {Riechert}}]{Consonni2009}%
  \BibitemOpen
  \bibfield  {author} {\bibinfo {author} {\bibfnamefont {V.}~\bibnamefont
  {Consonni}}, \bibinfo {author} {\bibfnamefont {M.}~\bibnamefont {Knelangen}},
  \bibinfo {author} {\bibfnamefont {U.}~\bibnamefont {Jahn}}, \bibinfo {author}
  {\bibfnamefont {A.}~\bibnamefont {Trampert}}, \bibinfo {author}
  {\bibfnamefont {L.}~\bibnamefont {Geelhaar}}, \ and\ \bibinfo {author}
  {\bibfnamefont {H.}~\bibnamefont {Riechert}},\ }\href@noop {} {\bibfield
  {journal} {\bibinfo  {journal} {Appl. Phys. Lett.}\ }\textbf {\bibinfo
  {volume} {95}},\ \bibinfo {pages} {241910} (\bibinfo {year}
  {2009})}\BibitemShut {NoStop}%
\bibitem [{\citenamefont {Grossklaus}\ \emph {et~al.}(2013)\citenamefont
  {Grossklaus}, \citenamefont {Banerjee}, \citenamefont {Jahangir},
  \citenamefont {Bhattacharya},\ and\ \citenamefont
  {Millunchick}}]{Grossklaus2013a}%
  \BibitemOpen
  \bibfield  {author} {\bibinfo {author} {\bibfnamefont {K.~A.}\ \bibnamefont
  {Grossklaus}}, \bibinfo {author} {\bibfnamefont {A.}~\bibnamefont
  {Banerjee}}, \bibinfo {author} {\bibfnamefont {S.}~\bibnamefont {Jahangir}},
  \bibinfo {author} {\bibfnamefont {P.}~\bibnamefont {Bhattacharya}}, \ and\
  \bibinfo {author} {\bibfnamefont {J.~M.}\ \bibnamefont {Millunchick}},\
  }\href@noop {} {\bibfield  {journal} {\bibinfo  {journal} {J. Cryst. Growth}\
  }\textbf {\bibinfo {volume} {371}},\ \bibinfo {pages} {142} (\bibinfo {year}
  {2013})}\BibitemShut {NoStop}%
\bibitem [{\citenamefont {Lang}(1974)}]{Lang1974}%
  \BibitemOpen
  \bibfield  {author} {\bibinfo {author} {\bibfnamefont {D.~V.}\ \bibnamefont
  {Lang}},\ }\href@noop {} {\bibfield  {journal} {\bibinfo  {journal} {J. Appl.
  Phys.}\ }\textbf {\bibinfo {volume} {45}},\ \bibinfo {pages} {3023} (\bibinfo
  {year} {1974})}\BibitemShut {NoStop}%
\bibitem [{\citenamefont {Sze}\ and\ \citenamefont {Ng}(2006)}]{SzeBook}%
  \BibitemOpen
  \bibfield  {author} {\bibinfo {author} {\bibfnamefont {S.~M.}\ \bibnamefont
  {Sze}}\ and\ \bibinfo {author} {\bibfnamefont {K.~K.}\ \bibnamefont {Ng}},\
  }\href@noop {} {\emph {\bibinfo {title} {Physics of Semiconductor Devices,
  3rd Edition}}}\ (\bibinfo  {publisher} {Wiley},\ \bibinfo {year}
  {2006})\BibitemShut {NoStop}%
\bibitem [{\citenamefont {Look}\ \emph {et~al.}(1996)\citenamefont {Look},
  \citenamefont {Reynolds}, \citenamefont {Kim}, \citenamefont {Aktas},
  \citenamefont {Botchkarev}, \citenamefont {Salvador},\ and\ \citenamefont
  {Morko\c{c}}}]{Look1996}%
  \BibitemOpen
  \bibfield  {author} {\bibinfo {author} {\bibfnamefont {D.~C.}\ \bibnamefont
  {Look}}, \bibinfo {author} {\bibfnamefont {D.~C.}\ \bibnamefont {Reynolds}},
  \bibinfo {author} {\bibfnamefont {W.}~\bibnamefont {Kim}}, \bibinfo {author}
  {\bibfnamefont {O.}~\bibnamefont {Aktas}}, \bibinfo {author} {\bibfnamefont
  {A.}~\bibnamefont {Botchkarev}}, \bibinfo {author} {\bibfnamefont
  {A.}~\bibnamefont {Salvador}}, \ and\ \bibinfo {author} {\bibfnamefont
  {H.}~\bibnamefont {Morko\c{c}}},\ }\href@noop {} {\bibfield  {journal}
  {\bibinfo  {journal} {J. Appl. Phys.}\ }\textbf {\bibinfo {volume} {80}},\
  \bibinfo {pages} {2960} (\bibinfo {year} {1996})}\BibitemShut {NoStop}%
\bibitem [{\citenamefont {Mott}(1969)}]{Mott1969}%
  \BibitemOpen
  \bibfield  {author} {\bibinfo {author} {\bibfnamefont {N.~F.}\ \bibnamefont
  {Mott}},\ }\href@noop {} {\bibfield  {journal} {\bibinfo  {journal} {Philos.
  Mag.}\ }\textbf {\bibinfo {volume} {19}},\ \bibinfo {pages} {835} (\bibinfo
  {year} {1969})}\BibitemShut {NoStop}%
\bibitem [{\citenamefont {Hill}(1971{\natexlab{a}})}]{Hill1971}%
  \BibitemOpen
  \bibfield  {author} {\bibinfo {author} {\bibfnamefont {R.~M.}\ \bibnamefont
  {Hill}},\ }\href@noop {} {\bibfield  {journal} {\bibinfo  {journal} {Philos.
  Mag.}\ }\textbf {\bibinfo {volume} {24}},\ \bibinfo {pages} {1307} (\bibinfo
  {year} {1971}{\natexlab{a}})}\BibitemShut {NoStop}%
\bibitem [{\citenamefont {Echeverr\'ia-Arrondo}, \citenamefont
  {P\'erez-Conde},\ and\ \citenamefont {Bhattacharjee}(2008)}]{Echeverria2008}%
  \BibitemOpen
  \bibfield  {author} {\bibinfo {author} {\bibfnamefont {C.}~\bibnamefont
  {Echeverr\'ia-Arrondo}}, \bibinfo {author} {\bibfnamefont {J.}~\bibnamefont
  {P\'erez-Conde}}, \ and\ \bibinfo {author} {\bibfnamefont {A.~K.}\
  \bibnamefont {Bhattacharjee}},\ }\href@noop {} {\bibfield  {journal}
  {\bibinfo  {journal} {J. Appl. Phys.}\ }\textbf {\bibinfo {volume} {104}},\
  \bibinfo {pages} {044308} (\bibinfo {year} {2008})}\BibitemShut {NoStop}%
\bibitem [{\citenamefont {Frenkel}(1938)}]{Frenkel1938}%
  \BibitemOpen
  \bibfield  {author} {\bibinfo {author} {\bibfnamefont {J.}~\bibnamefont
  {Frenkel}},\ }\href@noop {} {\bibfield  {journal} {\bibinfo  {journal} {Phys.
  Rev.}\ }\textbf {\bibinfo {volume} {54}},\ \bibinfo {pages} {647} (\bibinfo
  {year} {1938})}\BibitemShut {NoStop}%
\bibitem [{\citenamefont {Hill}(1971{\natexlab{b}})}]{Hill1971b}%
  \BibitemOpen
  \bibfield  {author} {\bibinfo {author} {\bibfnamefont {R.~M.}\ \bibnamefont
  {Hill}},\ }\href@noop {} {\bibfield  {journal} {\bibinfo  {journal} {Philos.
  Mag.}\ }\textbf {\bibinfo {volume} {23}},\ \bibinfo {pages} {59} (\bibinfo
  {year} {1971}{\natexlab{b}})}\BibitemShut {NoStop}%
\bibitem [{\citenamefont {Vincent}, \citenamefont {Chantre},\ and\
  \citenamefont {Bois}(1979)}]{Vincent1979}%
  \BibitemOpen
  \bibfield  {author} {\bibinfo {author} {\bibfnamefont {G.}~\bibnamefont
  {Vincent}}, \bibinfo {author} {\bibfnamefont {A.}~\bibnamefont {Chantre}}, \
  and\ \bibinfo {author} {\bibfnamefont {D.}~\bibnamefont {Bois}},\ }\href@noop
  {} {\bibfield  {journal} {\bibinfo  {journal} {J. Appl. Phys.}\ }\textbf
  {\bibinfo {volume} {50}},\ \bibinfo {pages} {5484} (\bibinfo {year}
  {1979})}\BibitemShut {NoStop}%
\bibitem [{\citenamefont {Jung}\ and\ \citenamefont {Kim}(2014)}]{Jung2014}%
  \BibitemOpen
  \bibfield  {author} {\bibinfo {author} {\bibfnamefont {E.}~\bibnamefont
  {Jung}}\ and\ \bibinfo {author} {\bibfnamefont {H.}~\bibnamefont {Kim}},\
  }\href@noop {} {\bibfield  {journal} {\bibinfo  {journal} {Phys. Status
  Solidi A}\ }\textbf {\bibinfo {volume} {211}},\ \bibinfo {pages} {1764}
  (\bibinfo {year} {2014})}\BibitemShut {NoStop}%
\bibitem [{\citenamefont {Mazzola}\ \emph {et~al.}(1994)\citenamefont
  {Mazzola}, \citenamefont {Saddow}, \citenamefont {Neudeck}, \citenamefont
  {Lakdawala},\ and\ \citenamefont {We}}]{Mazzola1994}%
  \BibitemOpen
  \bibfield  {author} {\bibinfo {author} {\bibfnamefont {M.~S.}\ \bibnamefont
  {Mazzola}}, \bibinfo {author} {\bibfnamefont {S.~E.}\ \bibnamefont {Saddow}},
  \bibinfo {author} {\bibfnamefont {P.~G.}\ \bibnamefont {Neudeck}}, \bibinfo
  {author} {\bibfnamefont {V.~K.}\ \bibnamefont {Lakdawala}}, \ and\ \bibinfo
  {author} {\bibfnamefont {S.}~\bibnamefont {We}},\ }\href@noop {} {\bibfield
  {journal} {\bibinfo  {journal} {Appl. Phys. Lett.}\ }\textbf {\bibinfo
  {volume} {64}},\ \bibinfo {pages} {2730} (\bibinfo {year}
  {1994})}\BibitemShut {NoStop}%
\bibitem [{\citenamefont {Shklovskii}\ and\ \citenamefont
  {Efros}()}]{Shklovskii-Book}%
  \BibitemOpen
  \bibfield  {author} {\bibinfo {author} {\bibfnamefont {B.~I.}\ \bibnamefont
  {Shklovskii}}\ and\ \bibinfo {author} {\bibfnamefont {A.~L.}\ \bibnamefont
  {Efros}},\ }\href@noop {} {\emph {\bibinfo {title} {Electronic Properties of
  Doped Semiconductors}}},\ Chap.~\bibinfo {chapter} {9}\BibitemShut {NoStop}%
\bibitem [{\citenamefont {Shockley}\ and\ \citenamefont
  {Read}(1952)}]{Shockley1952}%
  \BibitemOpen
  \bibfield  {author} {\bibinfo {author} {\bibfnamefont {W.}~\bibnamefont
  {Shockley}}\ and\ \bibinfo {author} {\bibfnamefont {W.}~\bibnamefont
  {Read}},\ }\href@noop {} {\bibfield  {journal} {\bibinfo  {journal} {Phys.
  Rev.}\ }\textbf {\bibinfo {volume} {87}},\ \bibinfo {pages} {835} (\bibinfo
  {year} {1952})}\BibitemShut {NoStop}%
\bibitem [{\citenamefont {Mattila}, \citenamefont {Seitsonen},\ and\
  \citenamefont {Nieminen}(1996)}]{Mattila1996}%
  \BibitemOpen
  \bibfield  {author} {\bibinfo {author} {\bibfnamefont {T.}~\bibnamefont
  {Mattila}}, \bibinfo {author} {\bibfnamefont {A.}~\bibnamefont {Seitsonen}},
  \ and\ \bibinfo {author} {\bibfnamefont {R.}~\bibnamefont {Nieminen}},\
  }\href@noop {} {\bibfield  {journal} {\bibinfo  {journal} {Phys. Rev. B}\
  }\textbf {\bibinfo {volume} {54}},\ \bibinfo {pages} {1474} (\bibinfo {year}
  {1996})}\BibitemShut {NoStop}%
\bibitem [{\citenamefont {Fritsch}, \citenamefont {Schmidt},\ and\
  \citenamefont {Grundmann}(2003)}]{Fritsch2003}%
  \BibitemOpen
  \bibfield  {author} {\bibinfo {author} {\bibfnamefont {D.~l.}\ \bibnamefont
  {Fritsch}}, \bibinfo {author} {\bibfnamefont {H.}~\bibnamefont {Schmidt}}, \
  and\ \bibinfo {author} {\bibfnamefont {M.}~\bibnamefont {Grundmann}},\
  }\href@noop {} {\bibfield  {journal} {\bibinfo  {journal} {Phys. Rev. B}\
  }\textbf {\bibinfo {volume} {67}},\ \bibinfo {pages} {1} (\bibinfo {year}
  {2003})}\BibitemShut {NoStop}%
\bibitem [{\citenamefont {Witowski}\ \emph {et~al.}(1999)\citenamefont
  {Witowski}, \citenamefont {Pakuła}, \citenamefont {Baranowski},
  \citenamefont {Sadowski},\ and\ \citenamefont {Wyder}}]{Witowski1999}%
  \BibitemOpen
  \bibfield  {author} {\bibinfo {author} {\bibfnamefont {A.~M.}\ \bibnamefont
  {Witowski}}, \bibinfo {author} {\bibfnamefont {K.}~\bibnamefont {Pakuła}},
  \bibinfo {author} {\bibfnamefont {J.~M.}\ \bibnamefont {Baranowski}},
  \bibinfo {author} {\bibfnamefont {M.~L.}\ \bibnamefont {Sadowski}}, \ and\
  \bibinfo {author} {\bibfnamefont {P.}~\bibnamefont {Wyder}},\ }\href@noop {}
  {\bibfield  {journal} {\bibinfo  {journal} {Appl. Phys. Lett.}\ }\textbf
  {\bibinfo {volume} {75}},\ \bibinfo {pages} {4154} (\bibinfo {year}
  {1999})}\BibitemShut {NoStop}%
\bibitem [{\citenamefont {Nguyen}\ \emph {et~al.}(2013)\citenamefont {Nguyen},
  \citenamefont {Zhang}, \citenamefont {Connie}, \citenamefont {Kibria},
  \citenamefont {Wang}, \citenamefont {Shih},\ and\ \citenamefont
  {Mi}}]{Nguyen2013}%
  \BibitemOpen
  \bibfield  {author} {\bibinfo {author} {\bibfnamefont {H.~P.~T.}\
  \bibnamefont {Nguyen}}, \bibinfo {author} {\bibfnamefont {S.}~\bibnamefont
  {Zhang}}, \bibinfo {author} {\bibfnamefont {A.~T.}\ \bibnamefont {Connie}},
  \bibinfo {author} {\bibfnamefont {M.~G.}\ \bibnamefont {Kibria}}, \bibinfo
  {author} {\bibfnamefont {Q.}~\bibnamefont {Wang}}, \bibinfo {author}
  {\bibfnamefont {I.}~\bibnamefont {Shih}}, \ and\ \bibinfo {author}
  {\bibfnamefont {Z.}~\bibnamefont {Mi}},\ }\href@noop {} {\bibfield  {journal}
  {\bibinfo  {journal} {Nano Lett.}\ }\textbf {\bibinfo {volume} {13}},\
  \bibinfo {pages} {5437} (\bibinfo {year} {2013})}\BibitemShut {NoStop}%
\bibitem [{\citenamefont {Kishino}\ and\ \citenamefont
  {Yamano}(2014)}]{Kishino2014a}%
  \BibitemOpen
  \bibfield  {author} {\bibinfo {author} {\bibfnamefont {K.}~\bibnamefont
  {Kishino}}\ and\ \bibinfo {author} {\bibfnamefont {K.}~\bibnamefont
  {Yamano}},\ }\href@noop {} {\bibfield  {journal} {\bibinfo  {journal} {IEEE
  J. Quantum Electron.}\ }\textbf {\bibinfo {volume} {50}},\ \bibinfo {pages}
  {538} (\bibinfo {year} {2014})}\BibitemShut {NoStop}%
\bibitem [{\citenamefont {Hierro}\ \emph {et~al.}(2002)\citenamefont {Hierro},
  \citenamefont {Arehart}, \citenamefont {Heying}, \citenamefont {Hansen},
  \citenamefont {Mishra}, \citenamefont {DenBaars}, \citenamefont {Speck},\
  and\ \citenamefont {Ringel}}]{Hierro2002}%
  \BibitemOpen
  \bibfield  {author} {\bibinfo {author} {\bibfnamefont {A.}~\bibnamefont
  {Hierro}}, \bibinfo {author} {\bibfnamefont {A.~R.}\ \bibnamefont {Arehart}},
  \bibinfo {author} {\bibfnamefont {B.}~\bibnamefont {Heying}}, \bibinfo
  {author} {\bibfnamefont {M.}~\bibnamefont {Hansen}}, \bibinfo {author}
  {\bibfnamefont {U.~K.}\ \bibnamefont {Mishra}}, \bibinfo {author}
  {\bibfnamefont {S.~P.}\ \bibnamefont {DenBaars}}, \bibinfo {author}
  {\bibfnamefont {J.~S.}\ \bibnamefont {Speck}}, \ and\ \bibinfo {author}
  {\bibfnamefont {S.~A.}\ \bibnamefont {Ringel}},\ }\href@noop {} {\bibfield
  {journal} {\bibinfo  {journal} {Appl. Phys. Lett.}\ }\textbf {\bibinfo
  {volume} {80}},\ \bibinfo {pages} {805} (\bibinfo {year} {2002})}\BibitemShut
  {NoStop}%
\bibitem [{\citenamefont {Tuomisto}\ \emph {et~al.}(2005)\citenamefont
  {Tuomisto}, \citenamefont {Saarinen}, \citenamefont {Lucznik}, \citenamefont
  {Grzegory}, \citenamefont {Teisseyre}, \citenamefont {Suski}, \citenamefont
  {Porowski}, \citenamefont {Hageman},\ and\ \citenamefont
  {Likonen}}]{Tuomisto2005}%
  \BibitemOpen
  \bibfield  {author} {\bibinfo {author} {\bibfnamefont {F.}~\bibnamefont
  {Tuomisto}}, \bibinfo {author} {\bibfnamefont {K.}~\bibnamefont {Saarinen}},
  \bibinfo {author} {\bibfnamefont {B.}~\bibnamefont {Lucznik}}, \bibinfo
  {author} {\bibfnamefont {I.}~\bibnamefont {Grzegory}}, \bibinfo {author}
  {\bibfnamefont {H.}~\bibnamefont {Teisseyre}}, \bibinfo {author}
  {\bibfnamefont {T.}~\bibnamefont {Suski}}, \bibinfo {author} {\bibfnamefont
  {S.}~\bibnamefont {Porowski}}, \bibinfo {author} {\bibfnamefont {P.~R.}\
  \bibnamefont {Hageman}}, \ and\ \bibinfo {author} {\bibfnamefont
  {J.}~\bibnamefont {Likonen}},\ }\href@noop {} {\bibfield  {journal} {\bibinfo
   {journal} {Appl. Phys. Lett.}\ }\textbf {\bibinfo {volume} {86}},\ \bibinfo
  {pages} {031915} (\bibinfo {year} {2005})}\BibitemShut {NoStop}%
\bibitem [{\citenamefont {Fern{\'{a}}ndez-Garrido}\ \emph
  {et~al.}(2015)\citenamefont {Fern{\'{a}}ndez-Garrido}, \citenamefont
  {L{\"{a}}hnemann}, \citenamefont {Hauswald}, \citenamefont {Korytov},
  \citenamefont {Albrecht}, \citenamefont {Ch{\`{e}}ze}, \citenamefont
  {Skierbiszewski},\ and\ \citenamefont {Brandt}}]{Fernandez-Garrido2015}%
  \BibitemOpen
  \bibfield  {author} {\bibinfo {author} {\bibfnamefont {S.}~\bibnamefont
  {Fern{\'{a}}ndez-Garrido}}, \bibinfo {author} {\bibfnamefont
  {J.}~\bibnamefont {L{\"{a}}hnemann}}, \bibinfo {author} {\bibfnamefont
  {C.}~\bibnamefont {Hauswald}}, \bibinfo {author} {\bibfnamefont
  {M.}~\bibnamefont {Korytov}}, \bibinfo {author} {\bibfnamefont
  {M.}~\bibnamefont {Albrecht}}, \bibinfo {author} {\bibfnamefont
  {C.}~\bibnamefont {Ch{\`{e}}ze}}, \bibinfo {author} {\bibfnamefont
  {C.}~\bibnamefont {Skierbiszewski}}, \ and\ \bibinfo {author} {\bibfnamefont
  {O.}~\bibnamefont {Brandt}},\ }\href@noop {} {\  (\bibinfo {year} {2015})},\
  \Eprint {http://arxiv.org/abs/1510.06512} {arXiv:1510.06512} \BibitemShut
  {NoStop}%
\bibitem [{\citenamefont {Zettler}\ \emph {et~al.}(2015)\citenamefont
  {Zettler}, \citenamefont {Hauswald}, \citenamefont {Corfdir}, \citenamefont
  {Musolino}, \citenamefont {Geelhaar}, \citenamefont {Riechert}, \citenamefont
  {Brandt},\ and\ \citenamefont {Fern\'{a}ndez-Garrido}}]{Zettler2015}%
  \BibitemOpen
  \bibfield  {author} {\bibinfo {author} {\bibfnamefont {J.~K.}\ \bibnamefont
  {Zettler}}, \bibinfo {author} {\bibfnamefont {C.}~\bibnamefont {Hauswald}},
  \bibinfo {author} {\bibfnamefont {P.}~\bibnamefont {Corfdir}}, \bibinfo
  {author} {\bibfnamefont {M.}~\bibnamefont {Musolino}}, \bibinfo {author}
  {\bibfnamefont {L.}~\bibnamefont {Geelhaar}}, \bibinfo {author}
  {\bibfnamefont {H.}~\bibnamefont {Riechert}}, \bibinfo {author}
  {\bibfnamefont {O.}~\bibnamefont {Brandt}}, \ and\ \bibinfo {author}
  {\bibfnamefont {S.}~\bibnamefont {Fern\'{a}ndez-Garrido}},\ }\href@noop {}
  {\bibfield  {journal} {\bibinfo  {journal} {Cryst. Growth Des.}\ }\textbf
  {\bibinfo {volume} {15}},\ \bibinfo {pages} {4104} (\bibinfo {year}
  {2015})}\BibitemShut {NoStop}%
\bibitem [{\citenamefont {Suzue}\ \emph {et~al.}(1996)\citenamefont {Suzue},
  \citenamefont {Mohammad}, \citenamefont {Fan}, \citenamefont {Kim},
  \citenamefont {Aktas}, \citenamefont {Botchkarev},\ and\ \citenamefont
  {Morko\c{c}}}]{Suzue1996}%
  \BibitemOpen
  \bibfield  {author} {\bibinfo {author} {\bibfnamefont {K.}~\bibnamefont
  {Suzue}}, \bibinfo {author} {\bibfnamefont {S.~N.}\ \bibnamefont {Mohammad}},
  \bibinfo {author} {\bibfnamefont {Z.~F.}\ \bibnamefont {Fan}}, \bibinfo
  {author} {\bibfnamefont {W.}~\bibnamefont {Kim}}, \bibinfo {author}
  {\bibfnamefont {O.}~\bibnamefont {Aktas}}, \bibinfo {author} {\bibfnamefont
  {a.~E.}\ \bibnamefont {Botchkarev}}, \ and\ \bibinfo {author} {\bibfnamefont
  {H.}~\bibnamefont {Morko\c{c}}},\ }\href@noop {} {\bibfield  {journal}
  {\bibinfo  {journal} {J. Appl. Phys.}\ }\textbf {\bibinfo {volume} {80}},\
  \bibinfo {pages} {4467} (\bibinfo {year} {1996})}\BibitemShut {NoStop}%
\end{thebibliography}%

\end{document}